\documentclass[twocolumn,preprint2]
{aastex631}
\usepackage{amsmath}
\usepackage[T1]{fontenc}

\def\lesssim{\mathrel{\hbox{\rlap{\hbox{\lower4pt\hbox{$\sim$}}}\hbox{$<$}}}}
\def\gtrsim{\mathrel{\hbox{\rlap{\hbox{\lower4pt\hbox{$\sim$}}}\hbox{$>$}}}}
\newcommand{\bea}{\begin{eqnarray}}
\newcommand{\eea}{\end{eqnarray}}

\usepackage{amsmath}

\shorttitle{Accretion onto rapidly spinning binary black holes}
\shortauthors{L. Combi et al}
\graphicspath{{./}{./figures/}}
\begin{document}

\title{Magnetized accretion onto rapidly spinning binary black holes:\\ mini-disk thermodynamics, magnetic transport, and dual jets}

\author[0000-0002-5427-1207]{Luciano Combi}
\affiliation{Perimeter Institute for Theoretical Physics, Waterloo, Ontario N2L 2Y5, Canada}
\affiliation{Kavli Institute for Particle Astrophysics and Cosmology, Stanford University, Stanford, CA 94305, USA}

\author[0000-0002-8659-6591]{Manuela Campanelli}
\affiliation{Center for Computational Relativity and Gravitation, Rochester Institute of Technology, 85 Lomb Memorial Drive, Rochester, NY 14623, USA}
\affiliation{School of Physics and Astronomy, Rochester Institute of Technology, 84 Lomb Memorial Drive, Rochester, NY 14623, USA}

\author[0000-0003-0220-5723]{Sean M. Ressler}
\affiliation{Department of Physics \& Astronomy, University of Tennessee, Knoxville, 1408 Circle Dr \#401, Knoxville, TN 37996, USA}
\affiliation{Canadian Institute for Theoretical Astrophysics, University of Toronto, 60 St. George Street, Toronto, ON M5S 3H8, Canada}

\author[0000-0001-6157-6722]{Alexander J. Dittmann \textsuperscript{*}}
\affiliation{School of Natural Sciences, Institute for Advanced Study, 1 Einstein Drive, Princeton, NJ 08540, USA}
\altaffiliation{NASA Einstein Fellow}

\author[0000-0002-3907-9583]{Federico Cattorini}
\affiliation{Dipartimento di Fisica ``G. Occhialini'', Universit\`a di Milano-Bicocca, Piazza della Scienza 3, I-20126 Milano, Italy}
\affiliation{INFN, Sezione di Milano-Bicocca, Piazza della Scienza 3, I-20126 Milano, Italy}
\affiliation{INAF, Osservatorio Astronomico di Brera, Via E. Bianchi 46, I-23807 Merate, Italy}

\begin{abstract} 
Supermassive binary black holes embedded in gas-rich environments are promising multi-messenger sources for pulsar timing arrays and future space-borne gravitational-wave interferometers. Their electromagnetic emission is governed by nonlinear plasma dynamics around the binary and is expected to inherit variability associated with the orbital motion thereof. Because active galactic nuclei are intrinsically stochastic, identifying robust binary signatures requires predictive models that connect large-scale circumbinary flows to the black holes. Previous relativistic simulations of accreting binaries have mostly focused on smaller separations and lower spins. We perform three-dimensional general relativistic magnetohydrodynamic simulations of a relaxed, magnetized circumbinary disk accreting onto an equal-mass binary black hole with a separation of 30 gravitational radii and dimensionless spin $\chi=0.9$. At these separations, the mini-disks around each black hole are persistent mass reservoirs but still show pronounced amplitude modulations governed by the eccentric circumbinary disk and the sloshing gas between the mini-disks. We analyze how the magnetic flux is transported from the circumbinary disk to the horizons, launching powerful dual jets with energy extraction efficiencies reaching $\simeq 40\%$. The horizon-threading flux and jet luminosity alternate between the two black holes, producing an on-off dual-jet state. The wide jet funnels interact above the binary and form a persistent current sheet favorable to reconnection. We explore the influence of mini-disk thermodynamics, comparing efficient and inefficient cooling inside the cavity. Hotter mini-disks are less massive, exhibit weaker coherent periodicity, and launch less luminous jets despite comparable horizon-threading magnetic flux.
\end{abstract}

\section{Introduction}

\subsection{Supermassive binary black holes as multi-messenger sources}

Nearly all massive galaxies host a supermassive black hole (SMBH) at their centers, which can power intense electromagnetic emission through gas accretion and relativistic outflows. Because galaxies frequently merge in hierarchical fashion, the formation of SMBH binaries should be a common outcome of cosmic evolution \citep{Begelman1980}. Dynamical friction, stellar scattering, and gas torques may drive these binaries from kiloparsec to subparsec scales, where gravitational-wave emission can eventually dominate their orbital evolution and drive them toward coalescence \citep{Merritt2004}. The efficiencies and relative importance of these processes throughout the evolution of SMBH binaries remain open questions \citep{Koss2019}.

The most massive binaries ($\gtrsim 10^8\,M_{\odot}$) emit gravitational waves in the nanohertz band once they reach separations of $\lesssim 10^{-3}\,{\rm pc}{\,(M/10^{8}\,M_{\odot}})^{1/3}$ \citep{Thorne1976}. Binaries with year-to-decade orbital periods are natural targets for pulsar timing arrays (PTA), which use the Earth–pulsar baselines as a Galactic-scale gravitational-wave detector \citep{Hellings1983}. More than 15 years of radio observations from different PTA collaborations world-wide have revealed evidence of a stochastic gravitational wave background \citep{Agazie2023, Agazie2023a}, broadly consistent with a population of inspiralling supermassive binaries. If this background is dominated by inspiralling SMBH binaries, amplitude and spectrum are correlated with the mass function of the present SMBH population \citep{Phinney2001}, offering a unique probe of their evolution. Recent analysis of the PTA data indicates that local estimates of the SMBH population under-predict the abundance and typical masses of BHs, while disfavoring a background dominated by only a few exceptionally massive systems \citep{SatoPolito2023}. Inspiral and merger of lower-mass massive binaries, on the other hand, are prime targets for space-borne interferometers such as LISA \citep{AmaroSeoane2012} in the millihertz band. These systems are expected to last many cycles within the band resulting in high signal-to-noise ratio detections and thus allowing precise measurements of their masses, spins, luminosity distances, and orbital dynamics.

SMBH binaries may become luminous, broad-band electromagnetic sources if enough gas is drawn toward the center of the newly-merged galaxy \citep{Bogdanovic2022, DOrazio2023}. Because a compact-binary phase only occupies a small fraction of the lifetime of an active galactic nuclei (AGN), observational campaigns require large samples and long temporal baselines \citep{Haiman2008, kelley2019Massive}. The search for supermassive binaries is thus particularly suitable for the flourishing era of large surveys, e.g., with the Vera Rubin Observatory's LSST, SDSS-V, and the planned near-infrared Roman Space Telescope. 

No individual SMBH binary has been unequivocally confirmed despite some promising candidates. Time-domain searches targeting year-long orbital periodicity, $P\simeq 1\,(M/10^8\,M_{\odot})\, {\rm yr}$, are one of the most direct discovery channels. Periodic signals may arise from relativistic Doppler boosting \citep{d2015relativistic}, self-lensing between the black holes \citep{ingram2021self,  Davelaar2021}, and non-linear modulations produced by the gas dynamics \citep{dascoli2018Electromagnetic,gutierrez2022electromagnetic}. Identifying orbital motion in photometric data, however, remains challenging given the highly stochastic variability and red noise of quasar light curves, which may result in false positives when only a few cycles are observed \citep{ElBadry2026}. Radio jet precession and large-scale morphological distortions provide a complementary and cleaner diagnostic \citep{abraham1999beaming, romero1999beaming, kiehlmann2025pks}, although these features are not unique to binaries and only a minority of AGNs are radio loud. These limitations motivate predictive accretion models that connect orbital dynamics near the binary to observable variability and relativistic outflows.

\subsection{Overview of binary accretion}

Accretion onto binaries differs drastically from accretion onto single gravitating bodies due to the strong quadrupolar, time-dependent potential that disrupts the structure of the disk \citep{Lubow1991,Artymowicz1994}. For rotationally supported flows, gravitational torques from binaries with mass-ratio $q=m_1/m_2 \lesssim 1$ carve a deep cavity in the disk of size $r_{\rm cav} \simeq2-3 \,r_{12}$, where $m_{A}$ is the mass of a black hole and $r_{12}$ is the binary separation. The binary excites eccentric modes and spiral waves that propagate in the circumbinary disk (CBD). From the edge of the CBD, thin streams fall into the binary in near-ballistic trajectories but are in part violently deflected back, shocking against the cavity wall. During this process, a portion of this material loses enough angular momentum to fall into the cavity and get captured by the binary \citep{Shi2015, Tiede2022}. 

Near the black holes, the captured gas can form rotationally-supported structures known as mini-disks. These mini-disks have an outer truncation radius given approximately by $r_{\rm t} \approx 0.35 \,r_{12}$, where the gravitational influence of the other body becomes of order unity \citep{Paczynski1977, bowen2017Relativistic}. On the other hand, the inner radius of the mini-disk is determined by the inner-most circular orbit (ISCO) radius, $r_{\rm ISCO}$, which is a strong function of spin. A mini-disk behaves as a persistent mass reservoir depending on the ratio of its inflow time, set by angular-momentum transport, to the characteristic timescale on which the streams replenish and modulate it. In addition to MHD turbulence, tidal forcing and stream-induced shocks can transport angular momentum through the mini-disks \citep{Ryan2017}. As the binary separation decreases, the distance between $r_{\rm ISCO}$ and $r_{\rm t}$ contracts, allowing the mini-disks to quickly process their supplied mass and exhibiting strong variability \citep{Gold2014, Paschalidis2021, Combi2022}. Mini-disks are ultimately disrupted at very small separations, although dissipation can continue within the cavity \citep{Ennoggi2025}.

A large corpus of work has investigated binary accretion using 2D Newtonian-viscous simulations, where detailed explorations of the parameter space and long-term evolutions are computationally feasible \citep[e.g.][]{MacFadyen2008, DOrazio2012, Farris2013, munoz2016pulsed, munoz2019hydrodynamics, Siwek2022, Dittmann2022, Dittmann:2023ztg, Dittmann:2025rtc}. Accurately capturing the dynamics near the black holes, as well as the MHD processes driving turbulent stresses and outflows require solving the three-dimensional general relativistic magnetohydrodynamics (GRMHD) equations with a time-dependent spacetime \citep[e.g.][]{farris2011Binary, Farris2012, Noble2012}. Such calculations are challenging because the dynamical timescales near the horizon are widely separated from the secular evolution of the outer disk.  One strategy is to first evolve the outer circumbinary flow while excising the central binary \citep{Noble2012, LopezArmengol2021} and then use the relaxed solution to initialize a horizon-resolving simulation \citep{Bowen2018, bowen2019Quasiperiodicity, Combi2022, Ennoggi2025, Ennoggi2025a}. This approach has enabled relativistic studies of equal-mass binaries accreting from geometrically thin CBDs at late-inspiral separations, while recent developments have extended such calculations through rapid inspiral and merger \citep{Ennoggi2025, Ennoggi2025a}. 
 
%In this work, we extend these studies to larger binary separations and rapidly-spinning black holes. This allows us to investigate quasi-steady state mini-disks as well as powerful interacting dual jets. We compare thermodynamic prescriptions representing efficiently cooled and hot mini-disks and examine how the resulting structure controls accretion variability, magnetic-flux accumulation, and relativistic outflows

\subsection{Summary of this work}

We investigate circumbinary accretion onto an equal-mass binary at a \textit{fixed} separation of $r_{12}= 30\,M$. Slowly-evolving binary black holes in the GW regime surrounded with gas are uniquely relevant as persistent EM precursors in the LISA band. At this wide separation (compared to previous GRMHD binary BH simulations) the mini-disks are more massive and stable in contrast to the rapidly depleting, transient mini-disks characteristic of tighter binaries ($r_{12} \lesssim 20M$).  This is the first horizon-resolving simulation of rapidly spinning binary black holes $(\chi=0.9)$ accreting from a circumbinary disk; see Fig.~\ref{fig:dualjets_3d} for a 3D rendering of the simulation. We also explore for the first time the dependence of the mini-disk properties on thermodynamic assumptions of the inner cavity, considering both hot and cold mini-disks accreting from a radiatively efficient, geometrically-thin circumbinary disk.

The paper is organized as follows. In Section 2, we present the numerical methods, including modeling of an approximate dynamical spacetime metric, quasi-relaxed initial data, and GRMHD evolution schemes. In Section 3 we present results of simulations of circumbinary thin-disk accretion. From Sections 3.1 to 3.5, we analyze variability and time-averaged properties of mini-disks. In Sections 3.6 and 3.7, we analyze the magnetic field properties in the circumbinary flows, how it is transported to mini-disks, and the behavior of the horizon-threading flux. In Section 3.8, we show the morphology and time-dependent properties of dual jets and their interaction. In Section 4 we present results comparing simulations of hot and cold mini-disk accretion. In Section 5 we compare with previous work and discuss caveats. We present our conclusions in Section 6.

\subsection{Conventions and notation}

We use the $(-+++)$ signature for the spacetime metric. We use geometric natural units where $G=c=1$, and Heaviside-Lorentz (HL) units for the magnetic field [related to Gaussian (G) units as $B^{\rm G}= \sqrt{4\pi} B^{\rm HL}$]. Barred coordinates, $\bar{r}^i=(\bar{x},\bar{y},\bar{z})$, denote the frames centered on  a BH; primed coordinates, $r'^i=(x',y',z')$, denote the frame corotating with the binary where the BHs are at rest at $y'=0$; and coordinates, $r^i=(x,y,z)$, denote global coordinates in the center of mass frame. 

We denote spherically-integrated, density-weighted quantities, time-averaged over a $\Delta t$ interval as $\langle \mathcal{Q} \rangle_{\rho}:= \Delta t^{-1}\int_{\Delta t} \lbrace \mathcal{Q \rho}\rbrace/ \lbrace \rho \rbrace dt$, where braces denote integration over a sphere, $\lbrace \mathcal{Q} \rbrace := \int \mathcal{Q} dA$, with the area element defined as $dA:= \sqrt{-g} d\phi d\theta$, and $g$ is the determinant of the metric.

\section{Numerical Methods}
\label{sec:numerical_methods}

\subsection{GRMHD evolution}
\label{subsec:grmhd_evolution}

We perform fully 3D simulations of the ideal general-relativistic magnetohydrodynamics (GRMHD) equations with a modified variant of the open-source, flux-conservative \texttt{GRHydro} code \citep{Mosta2014}, which is integrated into the Einstein Toolkit infrastructure \citep{Loffler2012}. The conservation laws are evolved in the Valencia formulation \citep{font2002Threedimensional}, employing enhanced PPM reconstruction \citep{reisswig2013ThreeDimensional} and the approximate Harten-van Lax-Leer Riemann solver \citep{harten1983Upstream} for the numerical fluxes. The magnetic-field evolution is carried out using a staggered vector potential in the generalized Lorenz gauge \citep{Etienne2010, Etienne2012}, combined with upwind constrained transport for the electric field \citep{DelZanna2007,  Mignone2021}. Recovery of primitive variables from the conserved quantities follows the scheme in \cite{siegel2018Recovery}. We use an ideal gas equation of state with adiabatic index $\Gamma=5/3$. We model photon cooling in the flow adding a source term to the conservation equations which is isotropic in the fluid frame and maintains the entropy of the gas \citep{Noble2012, bowen2017Relativistic}, maintaining, in turn, the initial geometrical thickness of the disk.

To ensure the stability of the MHD solution, particularly in magnetized regions such as the jet funnel, we impose several fixes to the primitives after recovery from conservatives. If rest-mass density $\rho<\rho_{\rm floor}=10^{-9}\,\rho_{\rm max}$, where $\rho_{\rm max}$ is the maximum density in the CBD, we inject mass to obtain $\rho=\rho_{\rm floor}$; if the magnetization $b^2/\rho>\sigma_{\rm max}=30$, we inject mass to obtain $\rho=b^2/\sigma_{\rm max}$, where $b$ is the comoving magnetic field density; if the Lorentz factor is larger than $\gamma_{\rm max}=10$,  we rescale all velocity primitives to ensure $\gamma=\gamma_{\rm max}$; if inverse plasma-beta is $b^2/2p>\beta^{-1}_{\rm max}=3000$, we inject specific internal energy until $\epsilon =b^2/[2\beta^{-1}_{\rm max}\rho(\Gamma-1)]$. Inside the black hole horizons, we additionally apply ceilings to $\rho$ and $\epsilon$ to ensure stability. Conservatives are recalculated if any of the above fix-ups are applied. Additionally, in the flux calculation, we obtain the fluxes using a Lax-Friedrich method when the solution is close to any of these floor values. 

We use zero-gradient (flat) boundary conditions for all hydrodynamical variables, $n^i\partial_i P=0$, enforcing outflow conditions for the velocity as $v^in_i=0$ if $v^in_i<0$, where $v^i$ is the primitive velocity and $n^i$ is a vector perpendicular to the boundary surface. For the vector potential, we impose $A_i n^i=0$ and interpolate the other two components linearly, which leads to $\epsilon_{ijk} B^i n^j=0$ (i.e. no tangential fields in the boundary face) and zero gradient for $B^in_i$ \citep{Mewes2020}. This condition prevents the outer boundary from supporting artificial
tangential magnetic stresses, surface currents, or anchored magnetic loops. In ideal MHD, a purely normal magnetic field carries no normal Poynting flux, so the boundary cannot inject spurious electromagnetic energy back into the domain. We found that a naive linear extrapolation of all components of $A_i$, to obtain a zero-gradient condition on $B^i$, can generate artificial Maxwell stresses and magnetic-field accumulation at the boundary, which can eventually drive unphysical inflows.

\begin{figure*}
        \centering
        \includegraphics[width=.9\linewidth]{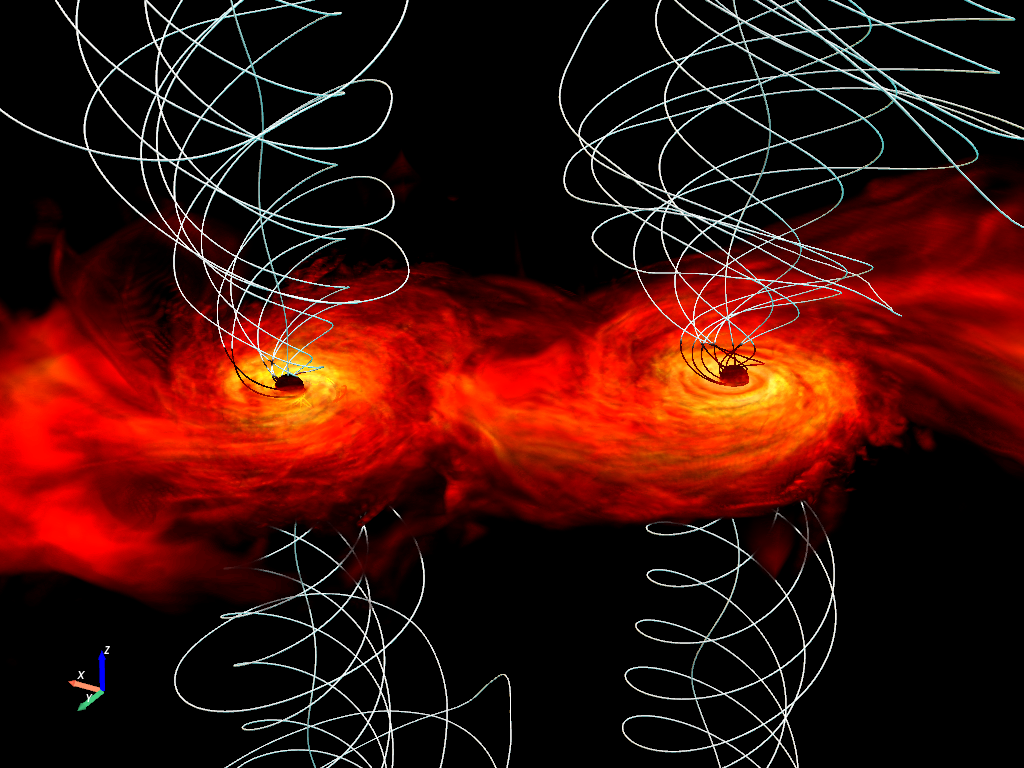} 
        \caption{Three-dimensional rendering of density and magnetic fields (white stream lines) showing the accreting mini-disks form around the spinning binary black hole and the transfer of mass between each other.}
        \label{fig:dualjets_3d}
\end{figure*}

\subsection{Spacetime}
\label{subsec:spacetime}

We model the spacetime using the superimposed Kerr-Schild approach \citep{Combi2026, Combi2021, LopezArmengol2021}, which is a general, strong-field approximation of a binary black hole metric. This method linearly superimposes a pair of time-dependent boosted Kerr-Schild metrics, which remains valid for all practical purposes even close to merger. It has been tested against other approximate binary metrics \citep{Combi2021} and numerical relativity \citep{Combi2026}, showing excellent accuracy for evolving MHD flows and substantial computational advantages over full numerical evolution of Einstein's equations. The black hole trajectories for the boosts are prescribed using a high-order post-Newtonian (PN) framework \citep{PN_ref1, PN_ref2}, capturing the orbital dynamics with great accuracy \citep{Csizmadia2012}.

We evolve a circular, equal-mass binary black hole at a fixed separation of $r_{12}=30\,M$ with aligned spins of $\chi^{z}_{A}=a^{z}_{A}/m_{A}=0.9$, where $m_{A}=0.5\,M$ is the mass of the black hole $A$ and $M=m_{1}+m_2$ is the total mass of the system. In solving the PN equations, we exclude radiative terms to fix the separation, but we include all other conservative corrections up to 4PN order; we use an eccentric reduction procedure to obtain a circular orbit. The orbital period is $P_{\rm bin} \approx 2\pi \,(30)^{3/2}\,M = 1030 \,M$, and the radius of the black hole horizon in these coordinates is $r_{\rm H} = (1+\sqrt{1-\chi_A^2})\,m_{A}=1.43\,m_{A}= 0.7 \,M.$

\subsection{Grid setup}

We use the \texttt{Carpet} driver \citep{Schnetter2004} in the Einstein Toolkit for Berger-Oliger, box-in-box mesh refinement with subcycling in time. The base grid is a cube of radius $1000\,M$ in each direction and resolution $dx_{\rm base}=8\,M$. We set mesh zones following each BH with $7$ refinement levels and inner-most radius of $9\,M$; a third mesh zone is centered at the center of mass of the binary with $6$ levels and inner radius of $60\,M$. The finer resolution is $dx_{\rm base}/2^7=\,M/16$, resolving the BH horizon radius by $\approx 12$ points. We use the method of lines for time evolution through fourth order Runge-Kutta with a CFL factor of $0.4$.

\subsection{Initial data: circumbinary disk equilibration}
\label{subsec:cbd_setup}

Circumbinary disks require many binary orbits to reach quasi-steady inflow equilibrium.
The local viscous time at the cavity edge where binary torques truncate the disk $(r_{\rm cav} \sim 2.5r_{12})$ is long,
$t_\nu(r_{\rm cav}) = [3\alpha\,\Omega(r_{\rm cav})]^{-1}(h/r)^{-2}$, 
which translates to
$t_{\nu}/P_{\rm bin} \simeq 0.2\,\alpha^{-1}(h/r)^{-2}$. Even for moderately thin, turbulent disks ($\alpha \sim 0.1$, 
$h/r \sim 0.1$) inflow equilibrium requires $t_{\nu}/P_{\rm bin}\gtrsim 200$ binary orbits, growing as $(h/r)^{-2}$ for radiatively-efficient flows. 
The cavity eccentricity and density modes likewise 
saturate on $\sim t_\nu(r_{\rm cav})$, consistent with the 
$\gtrsim \mathcal{O} (100\, P_{\rm bin})$ relaxation times measured in many published CBD simulations \citep[e.g.,][]{LopezArmengol2021,2020ApJ...889..114M, Dittmann2022}. Resolving these features is crucial to obtain the right accretion variability onto the black holes, which is controlled by the shape of the circumbinary disk.

Fully resolving horizon scales and evolving for hundreds of orbits is computationally expensive. We therefore evolve the circumbinary disk first with an excised interior for $\approx 200$ orbits, and we then port the solution as initial data to a simulation setup where the horizons are fully resolved. For equal-mass binaries, the presence of an accretion horizon \citep{Shi2012, Noble2012, Tiede2022} justifies excising the inner regions of the domain to focus computational resources on evolving the CBD. While this approach temporarily neglects feedback from black hole outflows and horizon-scale magnetic fields \citep{Most2024, Wang2025}, we assume it does not significantly alter the large-scale inflow solution, which is reasonable for the aligned spins and orbits that we use here.

We first evolve the binary-excised circumbinary disk using the spherical-coordinates based, ideal GRMHD code \texttt{SphericalNR} \citep{Mewes2020,Mewes2018}, which is adapted to the symmetries of the disk. We initially place a hydrodynamic torus of scale-height $h/r\approx0.1$ \citep{devilliers2003Magnetically, noble2009DIRECT} with a small magnetic loop seed with a density-weighted average $\beta \sim 100$. We evolve this initial configuration for $\approx 200$ binary orbits. This extended evolution allows the gravitational torques to carve out a stable, eccentric cavity that develops an $m=1$ density mode (lump) at the inner edge of the CBD, establishing a dynamically relaxed, steady mass supply to the inner cavity region \citep{noble2012Circumbinary, Shi2012}. This transition is achieved through a specialized ``hand-off'' procedure \citep{Ennoggi2025}. We interpolate the GRMHD primitive variables, i.e. the rest-mass density, pressure, vector potential, and fluid velocity from the spherical grid to the Cartesian AMR grid. Inside the cavity, the primitives are  extrapolated with an attenuation factor to reduce the transient effects and injected little energy in the initial data. Both codes evolve the magnetic vector potential, $A_i$, so the interpolation naturally maintains the divergence-free condition of the magnetic field. Following this hand-off, the evolution seamlessly resumes on the Cartesian grid, allowing us to capture the fully relativistic steady-state dynamics of the accreting spinning binary. There is, however, a short transient inside the cavity lasting $\simeq 5\,P_{\rm orb}$; we exclude this period from all time averages that we compute.

\section{Simulations of thin-disk accretion onto binary black holes}

\begin{figure*}
    \centering
    \includegraphics[width=1.0\linewidth]{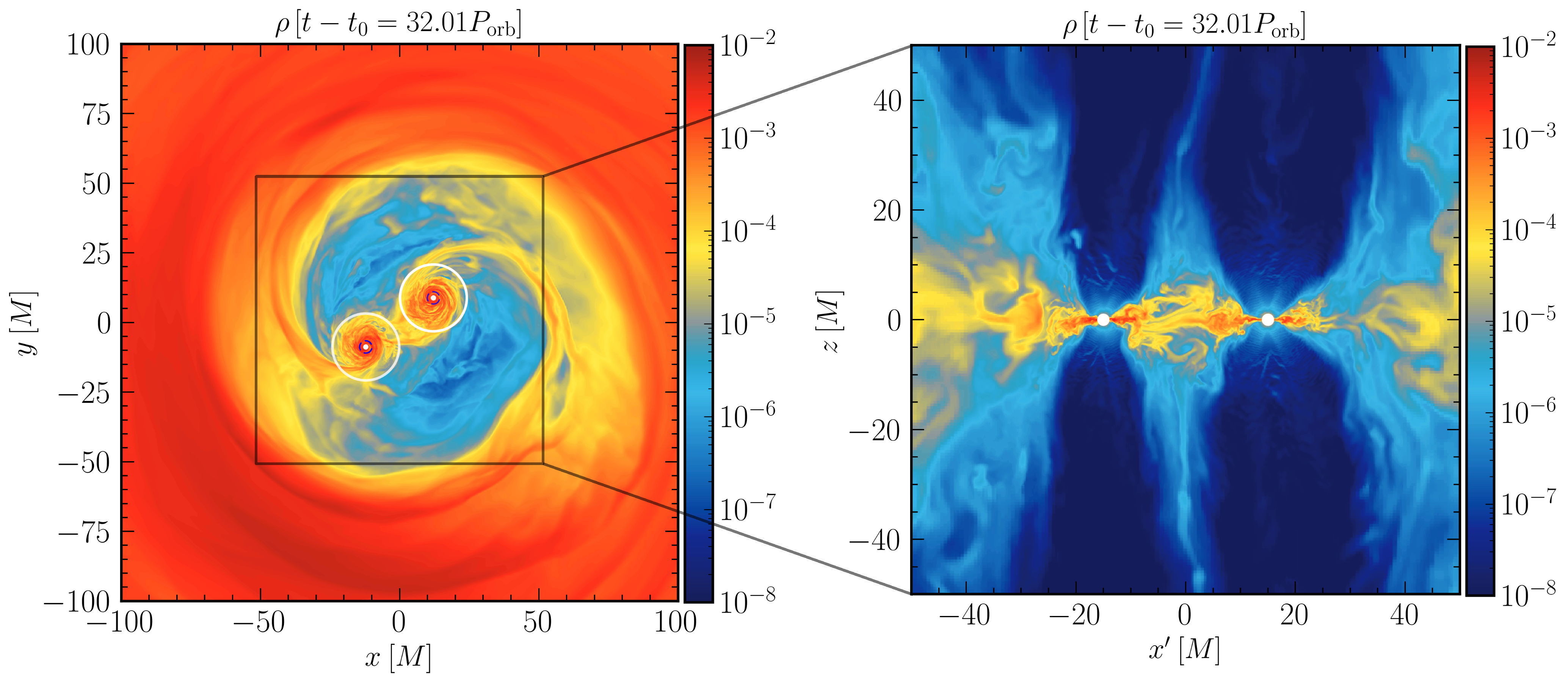}
    \caption{Rest-mass density, $\rho$, of the accretion structure in an equatorial slice (left) and a meridional slice (right) in the corotating frame of the binary. Mini-disks form inside the circumbinary cavity with radii approximately given by the Hill's sphere (white circles). The rapidly spinning black holes power dual Poynting-dominated jets which evacuate funnels.}
    \label{fig:rho_2d}
\end{figure*}

\subsection{Circumbinary disk and cavity morphology}

We show the global morphology of the accretion structure with equatorial (left) and meridional (right) snapshots of rest-mass density in Fig.~\ref{fig:rho_2d}.  The MRI-driven turbulent circumbinary disk is truncated due to strong binary torques. A strong one-arm ($m=1$) overdensity, known as ``the lump", appears near the eccentric inner edge of the disk, orbiting at the local Keplerian rate and supplying most of the mass to the binary when a BH passes nearby--- note that, due to the eccentric cavity, the companion black hole on the far side of the binary is too distant from the disk edge to accrete efficiently.
%(see Sec.~\ref{sec:variability} for further 
%discussion).

\begin{figure}
    \centering
    \includegraphics[width=1.0\linewidth]{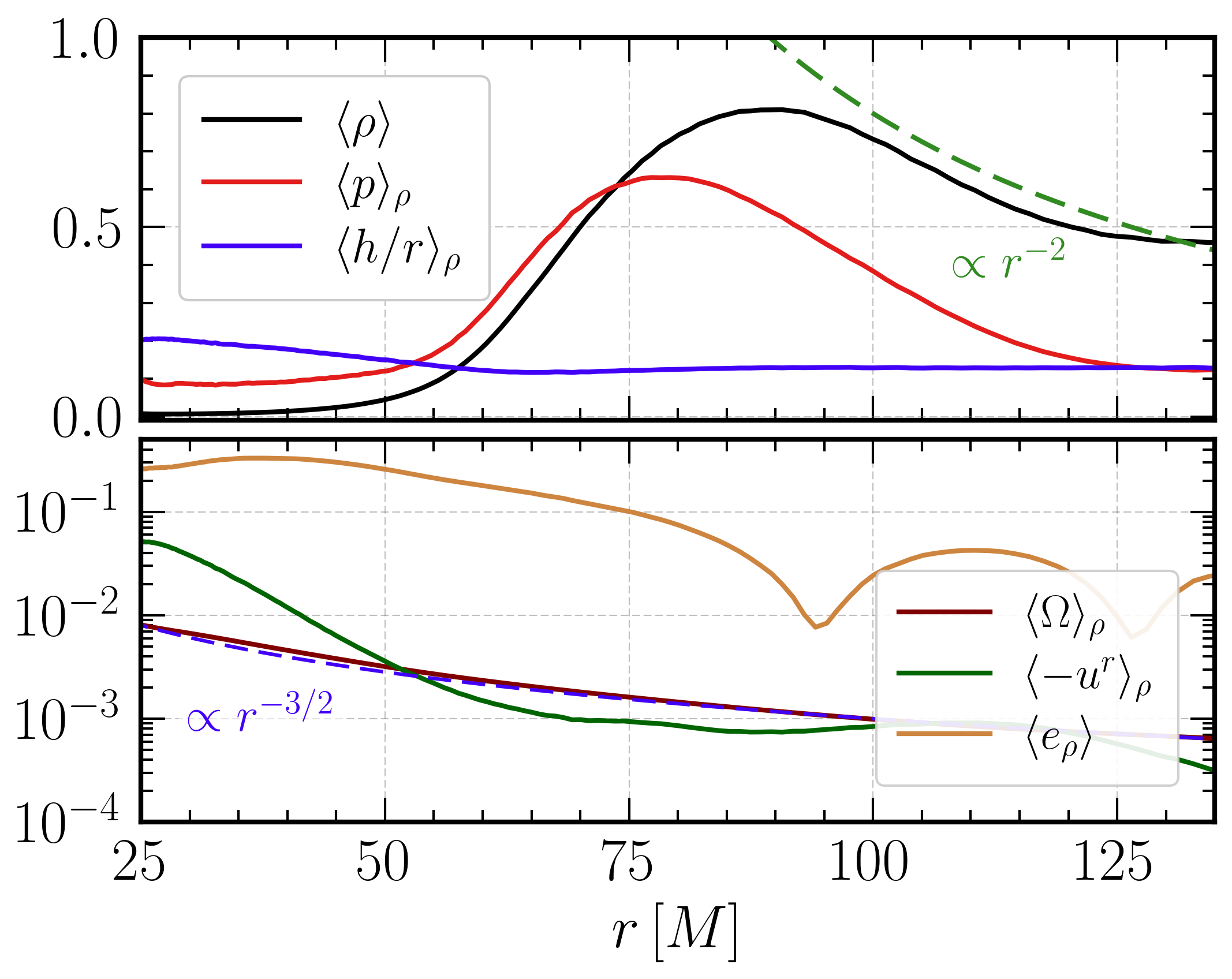}
    \caption{Spherical, time-averaged properties of the circumbinary disk as a function of radius. In top panel, we show density, density-weighted pressure, and scale-height ratio. In the bottom panel, we show density-weighted angular velocity, radial velocity, and eccentricity.}
    \label{fig:cbdravg}
\end{figure}

We show spherically-integrated, steady-state properties as a function of radius for the circumbinary disk in Fig.~\ref{fig:cbdravg}. The inner edge of the circumbinary disk, where the cavity starts, is defined at the maximum of density-weighted pressure, $\langle p \rangle_{\rho}$, which is located in our simulation at $r_{\rm cav}=75\,M\approx 2.5\,r_{12}$, consistent with previous simulations. Streams that fling away from the binary on high-angular-momentum trajectories shock against the disk at this location. 
%This pressure extremum is also a potential site for Rossby-wave instability \citep{Lovelace1999, MignonRisse2023}, which can act as a mass trapping  mechanism and give rise to the lump. 
We further define the time-average position of the lump as the peak density, located at $r\approx 90\,M$.
%The peak density at $r\approx 90\,M$ just outside the cavity wall determines the time-averaged position of the lump. 

%Beyond this radius, continuous injection of angular momentum 
%, mostly by $m=1$ spiral waves, 
%drives the disk into a decretion density profile $\rho\propto r^{-2}$ \citep{Pringle1991, rafikov2016Protoplanetary,Shi2015} .\footnote{For a steady disk with very small net mass-flux ($\dot{M}\approx0$) and 
%constant outward angular-momentum flux $F_J$, the viscous angular-momentum 
%equation reduces to $F_J = -2\pi r^3 \nu\Sigma\, d\Omega/dr$, which in the 
%Keplerian limit gives $\nu\Sigma \propto %r^{-1/2}$ 
%\citep[decretion-disk solution;][]{Pringle1991}. For an $\alpha$-disk with 
%$H/r\approx$~const, $\nu = \alpha c_s H \propto r^{1/2}$ and hence 
%$\Sigma \propto r^{-1}$, giving $\rho = \Sigma/H \propto r^{-2}$.}

\begin{figure*}
    \centering
    \includegraphics[width=1.0\linewidth]{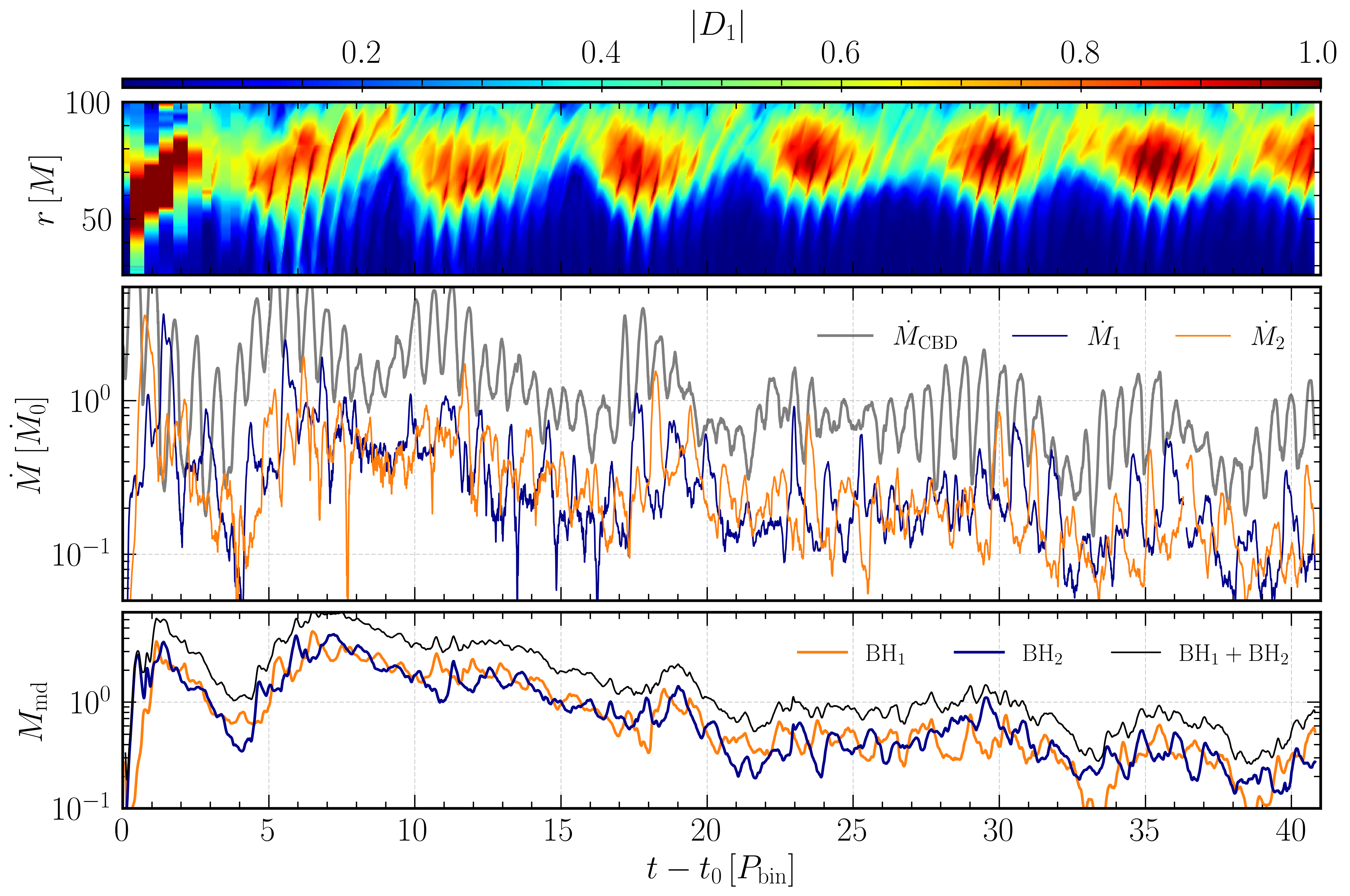}
    \caption{Time variability exhibited by the binary. (Top panel) Spherically-integrated azimuthal $m=1$ density mode as a function of radius and time, (middle panel) accretion rate onto each BH horizon and {into the} circumbinary cavity, and (bottom panel) mass contained inside the truncation radius of mini-disks around each black hole.}
    \label{fig:d1_mdot_mass}
\end{figure*}

The density inside the cavity is $\sim 10^{-3}$ times smaller than the disk, while the internal energy density $[u=p/(\Gamma-1)]$ is only $\sim 5$ times lower, so the \emph{specific} internal energy (and hence the temperature) is $\sim 200$ times higher in the cavity due to the continuous deposition 
of shock-heated gas. This is reflected in the gas scale height, which 
grows from the target $h/r\approx 0.1$ in the disk body to $h/r\approx 0.2$ 
in the cavity. This implies that heating occurs on a shorter timescale than the cooling time of $\sim 2\pi/\Omega$, consistent with persistent shocks from the near-ballistic streams.
%consistent with material falling ballistically in the cavity. 
The overall luminosity from the cavity is, however, lower due 
to the small density \citep{roedig2014OBSERVATIONAL}.
The bulk of the circumbinary disk has a moderate eccentricity of 
$e\sim 10^{-2}$, which grows significantly toward the cavity to reach $e\gtrsim 0.1$ in the edge. Inside the cavity, the rotation profile $\Omega(r)$ shown 
in Fig.~\ref{fig:cbdravg} departs from Keplerian while the fluid increases its radial velocity. To penetrate the cavity, the falling gas must have sufficiently low angular momentum to penetrate the centrifugal barrier of the binary; these streams fall from the edge of the cavity onto eccentric, almost ballistic orbits.

\subsection{Variability of mass fluxes}

Binary accretion is known to induce extreme hydrodynamical variability due to the time-dependent gravitational potential, especially for supersonic flows. This is evident in Fig.~\ref{fig:d1_mdot_mass}, where we show time series of accretion onto the BH horizons, $\dot{M}_A$, and through the circumbinary disk, $\dot{M}_{\rm CBD}$, mass contained in the mini-disk $M_{\rm md}$, and $D_1$, which measures the strength of the $m=1$ density modes as defined below. The accretion rate onto the horizon of an $A=1,2$ black hole is
\begin{equation}
    \dot{M}_{A} = -\int_{\bar{r}_H} \rho u^{\bar{r}}\, \sqrt{-g}\,d\bar{\phi} d\bar{\theta},
\end{equation}
where the area element $d\bar{A}$ and the radial velocity are defined in the comoving frame of the black hole; we normalize the accretion rate with $\dot{M}_0=5\times10^{-3}$. The mass contained in the mini-disk is defined as
\begin{equation}
    M_{\rm md,A} = \int^{r_{\rm t}}_{r_{\rm ISCO}} \rho \,d\bar{V},
\end{equation}
where $r_{t}=0.35 r_{12}$ is the truncation radius, and $r_{\rm ISCO}$ is the ISCO radius. The aziumuthal $m-$mode is defined in the center of mass frame as:
\begin{equation}
    D_{m}= \pi r^2\int \rho\,e^{-i m \phi} \, d\phi,
\end{equation}
where we use 2D data in the equatorial plane.

We observe a clear, long periodicity of $\sim 5\, P_{\rm bin}$ in all quantities, which corresponds to a peak in accretion from the CBD. This is a well known periodicity for equal-mass accreting binaries, associated with the motion of the lump at the edge of the cavity [$2\pi/\Omega(r_{\rm cav}) \approx 5 P_{\rm orb}$] observed in both MHD and hydrodynamical simulations \citep{MacFadyen2008, Shi2012, Noble2012, Lai2023}. The peak accretion event for this period happens when the lump passes through the pericenter of the cavity and dumps a lot of gas onto the binary. The $m=1$ mode evolution in the upper panel of Fig.~\ref{fig:d1_mdot_mass} shows that the relative power in density varies in time as it orbits around the cavity. After passing pericenter, the strength of the mode gets significantly reduced, $|D_1|\sim 0.5$, likely as a consequence of the accretion event; the lump then grows back again, reaches its peak, $|D_1|\sim 1$, at the next accretion event, and the cycle repeats. Radial variations of the lump with respect to the center of mass as shown in Fig.~\ref{fig:d1_mdot_mass} (upper panel) are due to the eccentricity of the cavity and will be important to explain short-timescale periodicity in the mini-disk.

\begin{figure}
    \centering
    \includegraphics[width=1.0\linewidth]{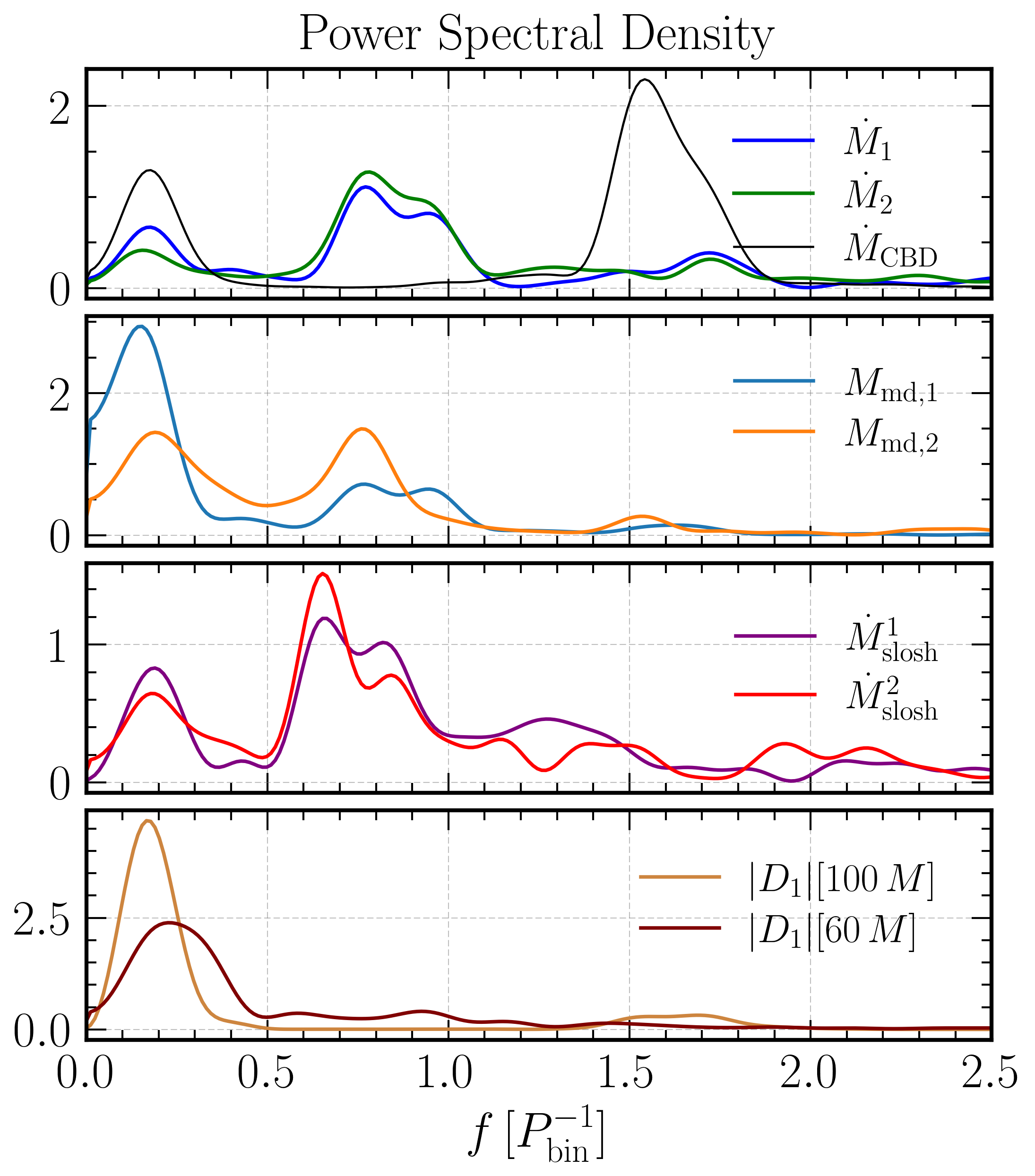}
    \caption{Power spectral densities (PSD) for BH accretion rate, mass in the mini-disk, sloshing mass transfer rate, and the $m=1$ azimuthal mode. These PSD were estimated using Welch’s method after subtracting a cubic polynomial trend from each time series. We adopted Hann-windowed segments of a common physical duration, with 50\% overlap and zero-padding to refine the sampled frequency grid. The spectra were normalized by their integrated power for visualization.}
    \label{fig:psd}
\end{figure}

The mini-disk masses and accretion rates present alternating periodicity throughout the evolution, with timescales of $\sim P_{\rm bin}$; variations in the amplitude of these quantities are accentuated (by a factor of $5-8$) every $5\,P_{\rm orb}$ during the larger accretion events from the circumbinary disk. To understand the origin of this short-scale periodicity we compute the power spectral density of mass fluxes, shown in Fig.~\ref{fig:psd}. The accretion rate and mini-disk mass show significant power at low frequencies, corresponding to the lump-driven accretion periodicity at $f_{\rm lump}\approx f_{\rm bin}/5$ described above. 

The higher-frequency {power} observed in the time series appears as a broad double-peak centered around $\approx 0.8 \, f_{\rm bin}$. This can be interpreted as a beat frequency between the motion of the lump and the binary frequency. Both mass and accretion rate {for an individual BH} increase periodically when {it} passes near the orbiting lump, {which happens at a beat frequency of} $f_{\rm beat} = f_{\rm bin} -f_{\rm lump}\approx 0.8\,f_{\rm bin}$. This feature has been identified in accreting equal-mass binaries simulations at small separations $r_{12}<20\,M$ \citep{Bowen2018, Combi2022, Avara2024}. The novel double peak feature at $f_{\rm beat}$ reflects the motion of the lump along the eccentric cavity: after an accretion event near pericenter, the binary continues to accrete when one of the holes aligns with the lump, but because the lump is slower through apocenter, the instantaneous beat frequency increases and splits the peak. This interpretation is supported by the presence of different peaks in the power spectrum of $|D_1|$ at $r=60\,M$ and $r=100\,M$ (bottom panel of Figure 5).  Note that these prominent peaks roughly correspond to the Keplerian frequencies of the lump. {Newtonian 2D simulations show larger, more eccentric cavities where the binary is too far from the apocenter to accrete from the lump and thus they do not show significant beat frequency variability{;} most show variability on the cavity/lump orbital period (for Mach numbers of $\sim 10$) with short-timescale variability depending on the gravitational softening and sink treatment \citep{Munoz2016, Dittmann2021, westernacher2022multiband}.}
%\ajdcom{Hrm. It seems like most Newtonian 2D sims don't really show *significant* beat frequency variability at any Mach number, while most show variability on the cavity/lump orbital period as well as the binary orbital period... However, at higher Mach numbers (20+) they rarely seem to show much of a `lump' but more so something like a high-density ridge along the inner cavity...} \luc{Sorry there was a typo, I meant to say that they do not see beat frequency, so I agree. Interesting point about those higher Mach cases. Also, I wonder how much MHD changes the cavity eccentricity.}

\begin{figure*}
    \centering
    \includegraphics[width=0.48\linewidth]{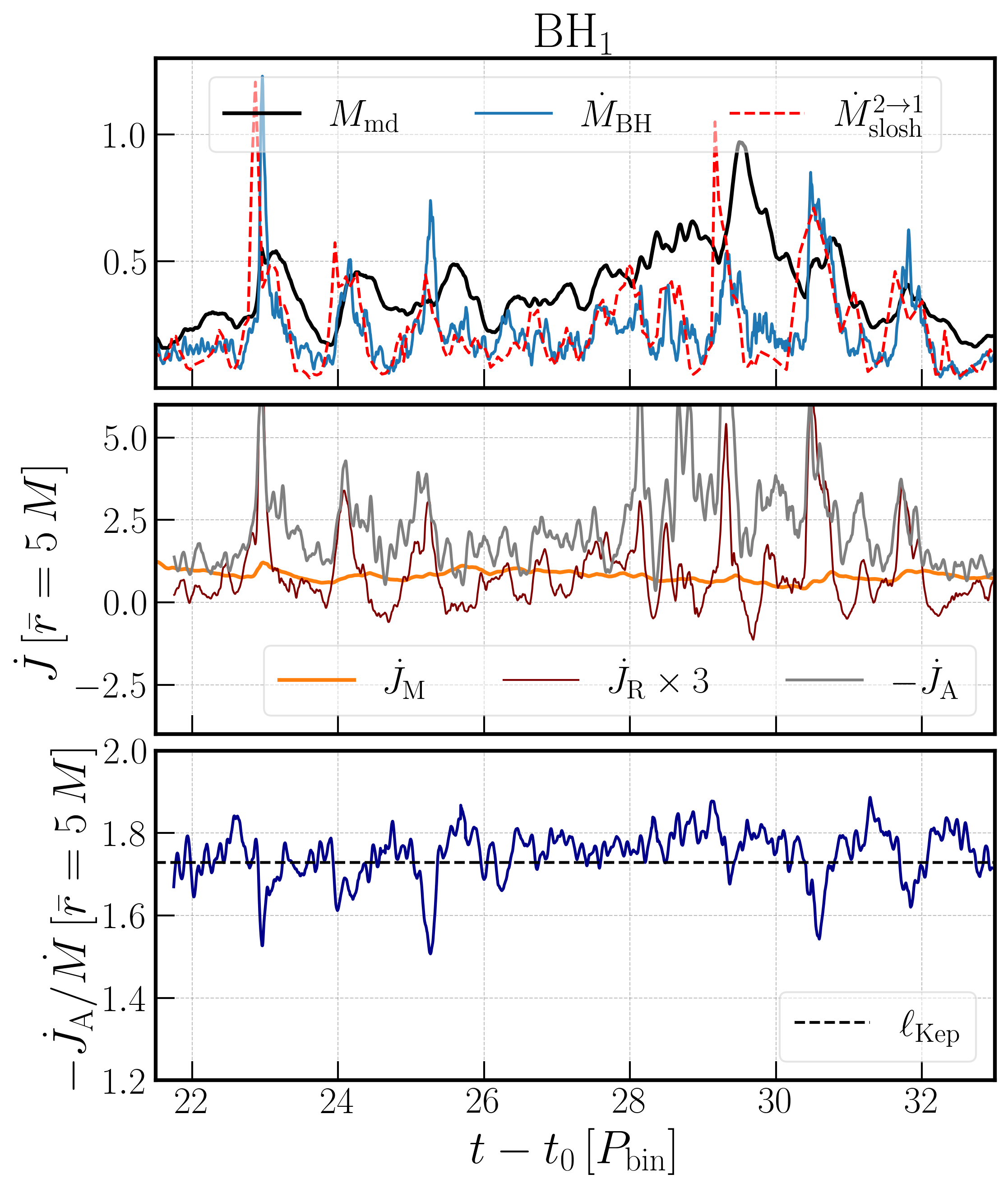}
    \includegraphics[width=0.48\linewidth]{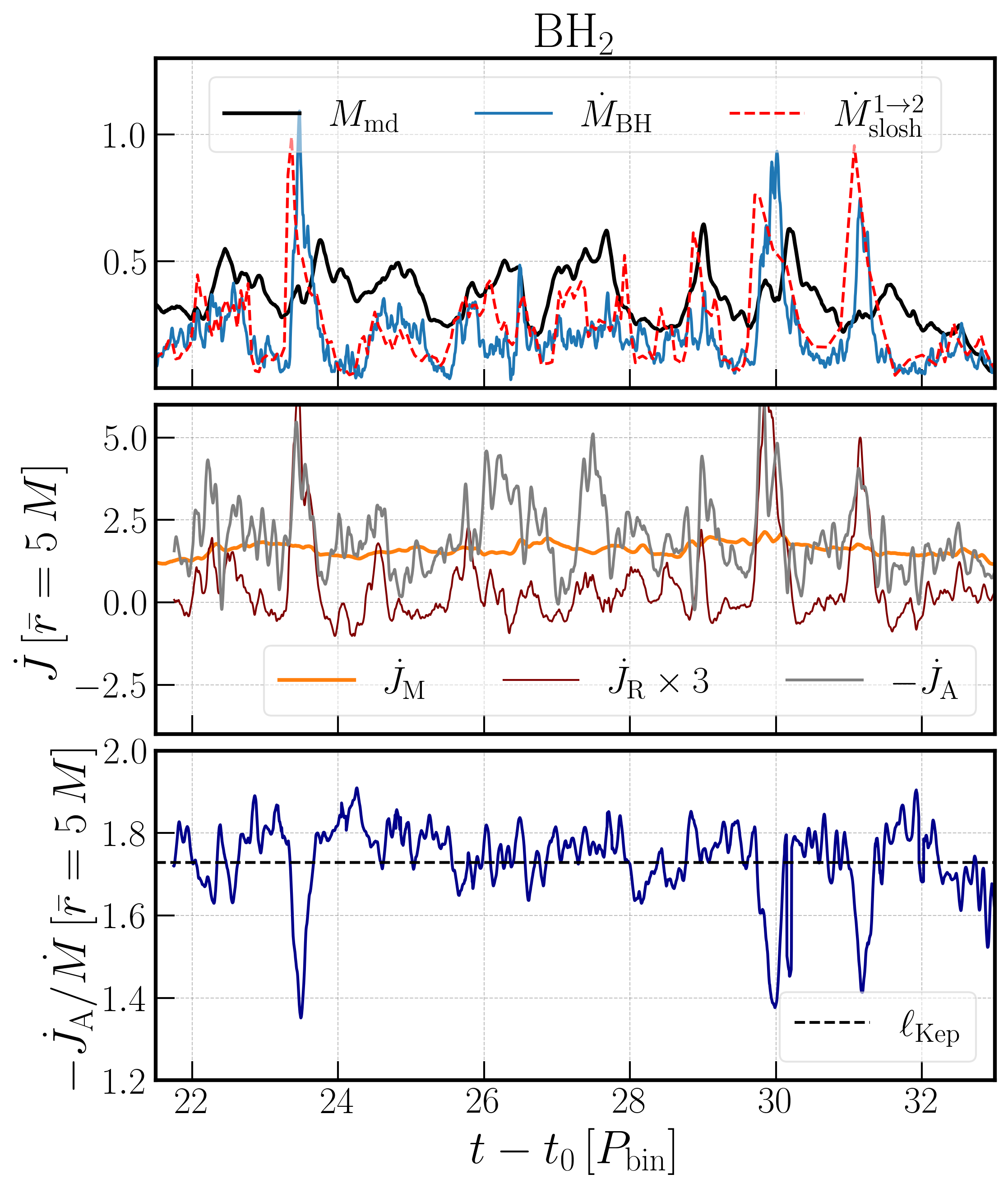}
    \caption{(Top panel) Detrended time evolution of mini-disk mass, mass flux onto the horizon, and mass flux crossing from one mini-disk to the other for BH$_1$, left panel, and BH$_2$, right panel. (Middle panel) Angular momentum fluxes through the mini-disk at $\bar{r}=5\,M$ in the frame of the BH, measured at a radius $\bar{r}=5\,M$. We divide the fluxes into Maxwell ($\dot{J}_{M}$), Reynolds ($\dot{J}_R$), and advective contributions ($\dot{J}_{\rm A}$), which we normalize with $10^{-3}$.  (Bottom panel) Specific angular momentum advected through the mini-disk with dashed line representing the geodesic circular specific angular momentum at that radius.}
    \label{fig:mdot_mass_slush}
\end{figure*}

\subsection{Mass exchange between mini-disks}

Gas captured from the circumbinary streams circularize within each mini-disk. Angular-momentum redistribution expands the outer layers toward the tidal
truncation radius where gas can cross the inner Roche-lobe boundary toward the companion. The dynamical importance of this mass exchange that we call \textit{sloshing} was first identified in relativistic hydrodynamic simulations and becomes increasingly prominent as the binary separation decreases \citep{bowen2017Relativistic}.

\begin{figure*}[ht!]
        \centering
        \includegraphics[width=0.48\linewidth]{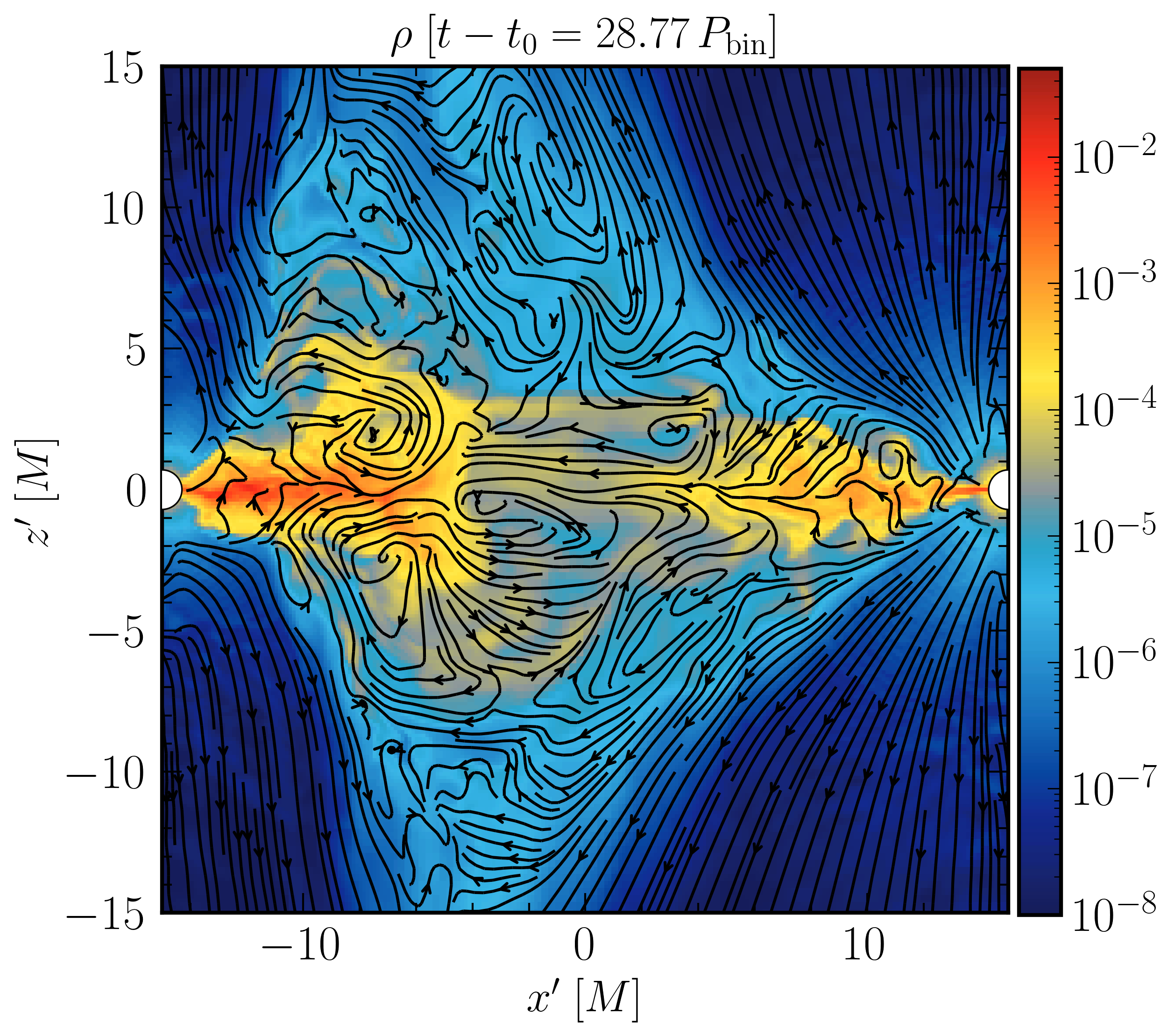}
        \includegraphics[width=0.48\linewidth]{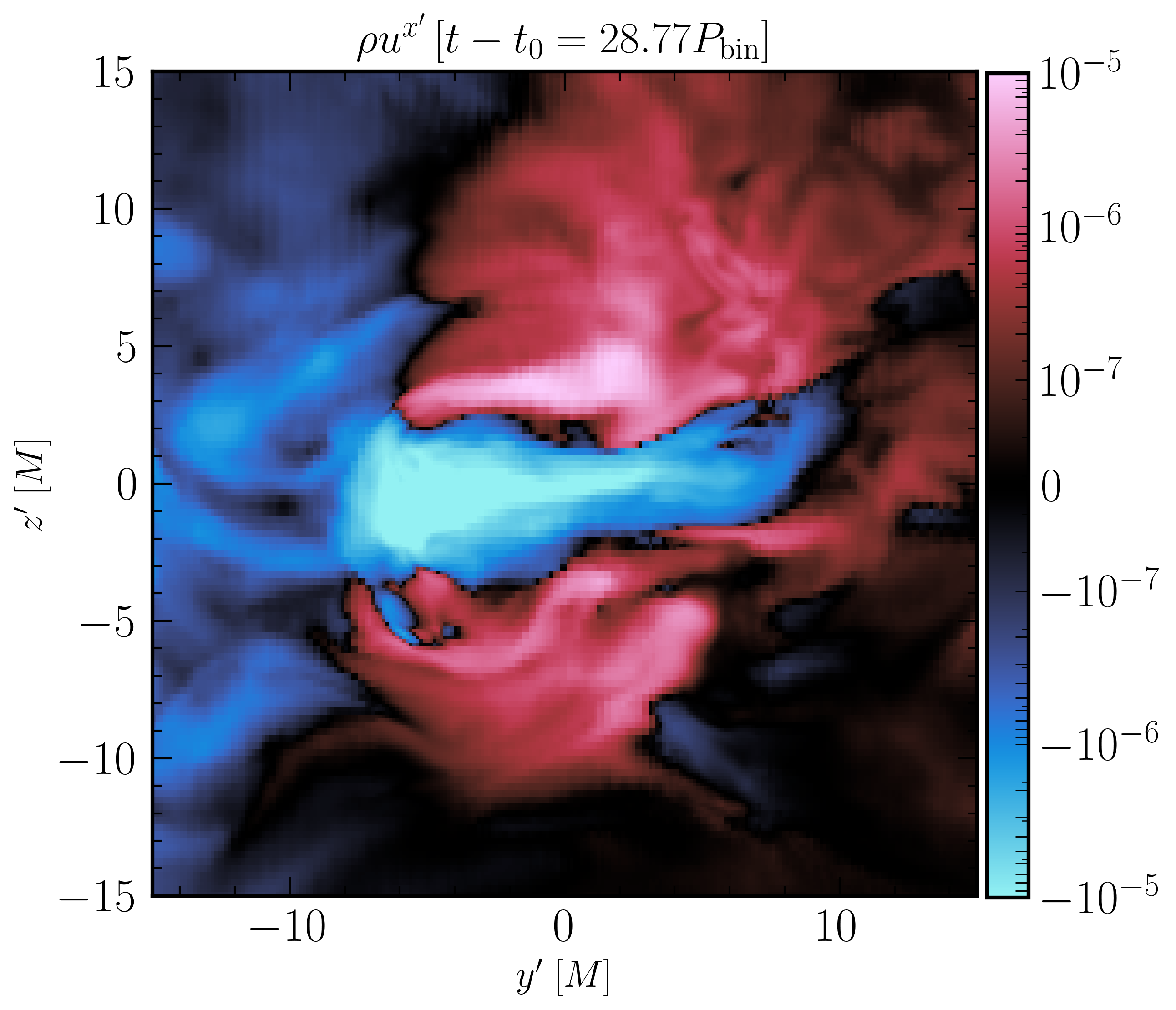}
        \caption{Mass transfer between black holes. (Left) Meridional plot in the corotating frame showing rest-mass density juxtaposed with coordinate transport velocity $(\alpha v^{i'}-\beta^{i'})$ field lines. (Right) Meridional plot orthogonal to the corotating plane showing mass flux density $\rho u^{x'}$ of the gas flowing from one mini-disk to the other.}
        \label{fig:sloshing}
\end{figure*}

We measure mass flux exchange between mini-disks through the plane $x'=0$, perpendicular to the binary separation vector; in these corotating coordinates, BH$_1$ is at $x'=15\,M$ and BH$_2$ at $x'=-15\,M$ (see Fig.~\ref{fig:sloshing}). We thus define the positive fluxes
\begin{align}
    \dot{M}_{\rm slosh}^{2\rightarrow1}
    &=
    \int_{\mathcal S} dy'\,dz'\,\sqrt{-g'}\,
    \rho u^{x'}\Theta(u^{x'}),\\
    \dot{M}_{\rm slosh}^{1\rightarrow2}
    &=
    -\int_{\mathcal S} dy'\,dz'\,\sqrt{-g'}\,
    \rho u^{x'}\Theta(-u^{x'}),
\end{align}
where $\mathcal S$ is restricted to $|y'|,|z'|<15\,M$, $u^{x'}$ is the four-velocity component in the corotating coordinates, and $\Theta$ is the Heaviside function. 

The sloshing fluxes have a strong variability and occur in quasi-periodic bursts, as shown in the zoom-in time series in Fig.~\ref{fig:mdot_mass_slush}; these bursts are stronger during the lump accretion events when more mass is brought to the cavity, around $\simeq23\,P_{\rm bin}$ and $\simeq 30\,P_{\rm bin}$ in this plot. The variability of $\dot{M}_{\rm slosh}$ is tightly correlated with the accretion rate onto the black holes, $\dot{M}_{\rm BH}$. The sloshing flux and the accretion rate onto the horizon share the same dominant beat frequency and have similar peak amplitude, growing by a factor of $\sim 5$, separated only by a short time delay (Fig.~\ref{fig:psd}). This close correspondence indicates
that a large fraction of the gas transferred during each sloshing event reaches the receiving black hole.

The associated horizon-accretion bursts coincide with sharp decreases in the angular momentum advected per unit accreted mass at $\bar r=5\,M$, $\ell_{\rm in}=-J_{\rm A}/\dot{M}$ (Fig.~\ref{fig:mdot_mass_slush}, bottom panels). The transferred gas therefore reaches the inner mini-disk with comparatively low angular momentum, allowing it to plunge rapidly toward the horizon rather than joining the rotationally supported reservoir. The sloshing burst occurs as the mini-disk starts being replenished by gas from the circumbinary disk, and decays before the mini-disk mass reaches a local maximum, as shown by black curves in upper panels of Fig.~\ref{fig:mdot_mass_slush}. The temporal ordering indicates that the accretion burst is associated with the inter mini-disk gas exchange, rather than with the draining of the mini-disk reservoir.

At larger binary separations, where the mini-disks are larger
and the exchanged mass is expected to be smaller relative to the mini-disk
reservoir \citep{bowen2017Relativistic}, this mechanism may become less
pronounced. Nevertheless, tidally driven shocks may still provide efficient
angular-momentum transport and generate variability on timescales of order
$P_{\rm bin}$ \citep{Ryan2017, Munoz2016, Rafikov2016}.

The three-dimensional structure of the sloshing flow is illustrated in
Fig.~\ref{fig:sloshing}. The left panel shows streamlines of the
corotating coordinate transport velocity,
$u^{i'}/u^{t'}=\alpha v^{i'}-\beta^{i'}$, projected onto the meridional
$(x',z')$ plane. In the displayed snapshot, gas is transferred
predominantly from the right mini-disk toward the left mini-disk. The
right panel shows the mass flux $\rho u^{x'}$ through the plane $x'=0$. The dominant transfer
occurs through a single equatorial stream centered near
$y'\simeq-5\,M$, where $\rho u^{x'}<0$ (blue contours) indicates motion toward the left black hole. A weaker counterflow with
$\rho u^{x'}>0$ (pink contours) occurs primarily at higher latitudes.

The inter-mini-disk  flows interact near the binary center of mass, where their collision produces vertically directed plumes. These are visible as the convergence and subsequent vertical deflection of the streamlines above and below the midplane in the left panel of Fig.~\ref{fig:sloshing}. A fraction of this shock-heated material becomes unbound and escapes along the low-density region between the two jet funnels. Its vertical collimation appears to be aided by lateral confinement from the surrounding dual jets.

\subsection{Angular momentum transport and accretion cycles}

The variability of mini-disk mass{es} and accretion rate{s} onto the black holes is controlled by several processes such as (i) mass supply from the circumbinary disk, (ii) angular-momentum transport and inflow through the mini-disk, (iii) gas exchange from the companion mini-disk, and (iv) mass loss in outflows.  As discussed in previous sections, accretion from the circumbinary disk is modulated by the orbital motion of the lump in our setup. The higher-frequency variability we observe in the BH accretion rate is instead governed by how rapidly the mini-disk processes the supplied mass, which, in turn, depends on the mechanism{s} transporting angular momentum inside the truncation radius. It is useful to split the total MHD angular momentum flux computed in the BH frame as
\begin{equation}
    \dot{J} = \lbrace T^{\bar{r}}_{\bar{\phi}} \rbrace = \dot{J}_{\rm M} + \dot{J}_{\rm A} +\dot{J}_{\rm R}, 
\end{equation}
where $\dot{J}_{\rm M}:= \lbrace b^2 u^{\bar{r}} u_{\bar{\phi}}- b^{\bar{r}} b_{\bar{\phi}} \rbrace$ is the total electromagnetic angular momentum flux, $\dot{J}_{\rm A}:=\lbrace \rho h u^{\bar{r}} \rbrace \lbrace \rho h u_{\bar{\phi}} \rbrace/\lbrace \rho h \rbrace $ is the advection component of the flux, and $\dot{J}_{\rm R}:= \lbrace \rho h u^{\bar{r}}u_{\bar{\phi}} \rbrace - \dot{J}_{\rm A}$ is the Reynolds (turbulent) component.

For comparison, the inflow time of a steady-state mini-disk extending from the ISCO to the truncation radius can be estimated as $t_{\rm infl}/P_{\rm bin} \sim [40\,\alpha\, (h/r_{\rm t})^2]^{-1} (1-\sqrt{\bar{r}_{\rm ISCO}/\bar{r}_{t}}) \lesssim 25$, for $\alpha,h/\bar{r}\sim0.1$. This estimate depends most 
sensitively on the mini-disk thermodynamics and dissipation but is nearly independent of separation, except when $r_t$ approaches $r_{\rm ISCO}$ ($r_{12} \lesssim 20\,M$). Accretion variability on timescales of  order $P_{\rm bin}$ -- much shorter than the viscous inflow time -- can nonetheless be imprinted on the BH accretion rate if the mass supplied per cycle from the CBD is comparable to the steady-state mini-disk reservoir, when shock-mediated angular momentum transport bypasses the viscous timescale, or if gas plunges onto the black holes directly with low-angular momentum. The first condition is likely satisfied  for the lump-driven periodicity at $\sim 5\,P_{\rm bin}$, as observed here and hinted at in previous scale-free two-dimensional Newtonian circumbinary-disk simulations. The
shorter-period beat modulation of order $\sim 1.25\,P_{\rm bin}$ in the accretion rate is in our case triggered by the sloshing of gas between the mini-disks, which brings low-angular momentum gas from the companion.

\begin{figure}
    \centering
    \includegraphics[width=1.0\linewidth]{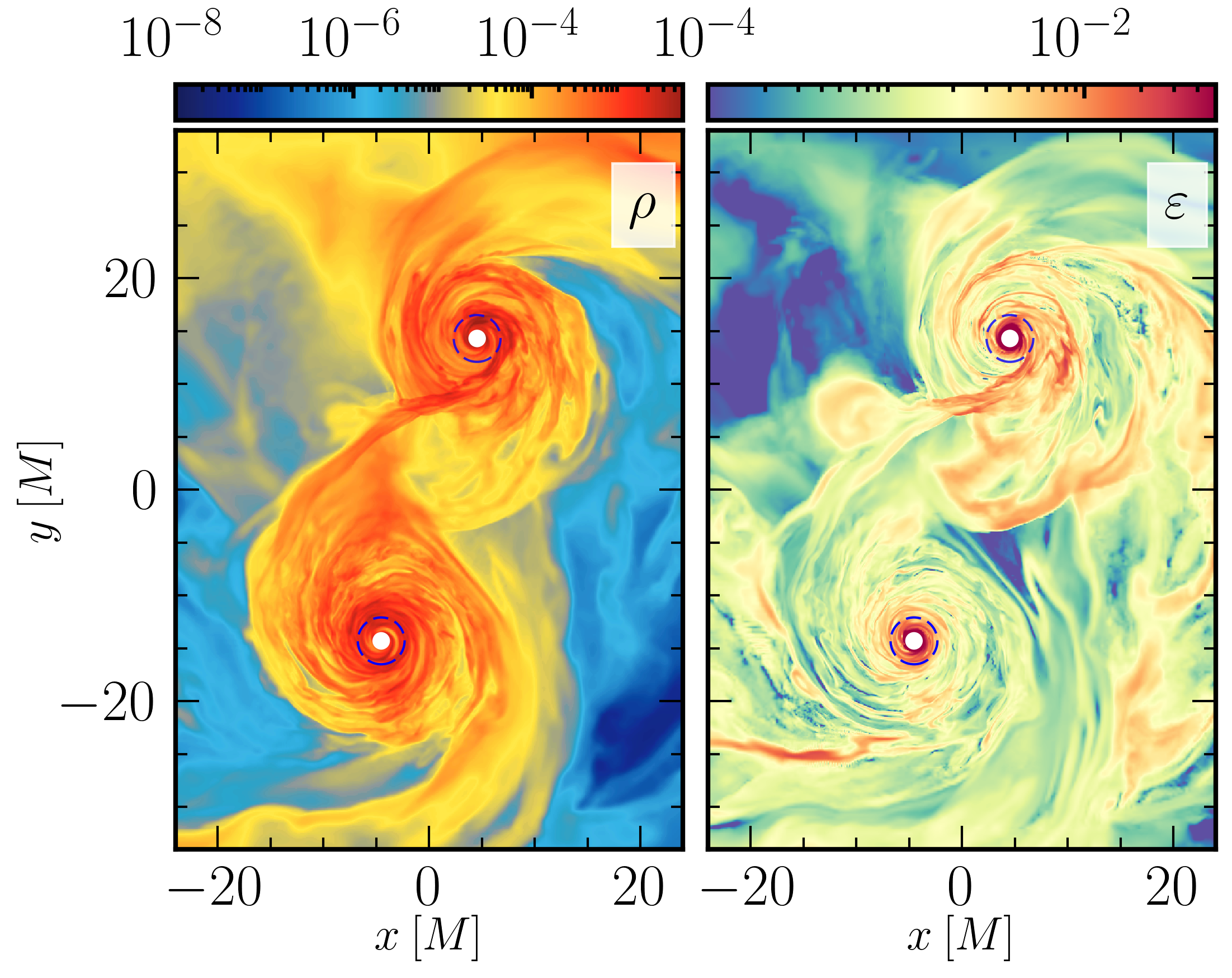}
    \caption{Density (left) and specific internal energy (right) in the equatorial plane during a mass sloshing event, where the mini-disk in the bottom is transferring the excess gas to the top mini-disk, heating it in the process. Dashed line circle represents the ISCO and the white disk represents the black hole horizon.}
    \label{fig:rhoeps_2d}
\end{figure}

We now summarize the recurrent cycle that organizes the accretion onto each black hole, analyzing the different contributions of angular momentum fluxes with respect to the BH frame. Each cycle
proceeds in roughly five phases.
\begin{itemize}

\item[1] \emph{Slosh impact}: as BH$_1$ approaches the lump, a stream overflowing from mini-disk~2 reaches mini-disk~1 first and
shocks against its outer edge.

\item[2]  \emph{Accretion burst}: The impact deposits low-angular-momentum gas, drives a transient enhancement of the
Reynolds stress $\dot{J}_R$, and triggers a prompt spike in the horizon accretion rate $\dot{M}_{1}$
together with a large negative excursion in the advective angular momentum flux $\dot{J}_{\rm A}$.

\item[3] \emph{CBD loading}: The mini-disk mass $M_{\rm md,1}$ starts growing from the lump-driven CBD stream and keeps rising after the accretion rate burst drops, indicating that the accreted burst is dominated by (shocked) gas plunging on a near-dynamical timescale rather than by viscous draining of the reservoir.

\item[4]  \emph{Relaxation}: with the CBD mass supply diminishing, mini-disk~1 settles into a quasi-steady state with $M_{\rm md,1}$ slowly declining and $\dot{M}_{1}$ roughly constant. Angular momentum transport by Maxwell stress, $\dot{J}_M$, becomes dynamically important with still large contributions from advective fluxes, $|\dot{J}_M|\sim |\dot{J}_{\rm A}|$.

\item[5] \emph{Reverse slosh}: a fraction of the residual mini-disk mass overflows toward mini-disk~2 once the gas spreads to the truncation radius, initiating the next cycle on the companion (Fig.~\ref{fig:rhoeps_2d}). The same sequence operates on BH$_2$ with a phase offset set by the binary's orbital phase relative to the lump, so that the two mini-disks alternate between supply-driven and relaxed states.

\end{itemize}

The bursts in horizon accretion and advective angular-momentum flux are thus
initiated by inter-mini-disk sloshing, while the circumbinary disk provides the dominant mass reservoir, most prominently on the lump-modulated
timescale. Maxwell stresses become relevant to the angular-momentum budget primarily during the relaxation phase between sloshing events. This cycle and the importance of the sloshing gas are in close agreement with the findings of \cite{Avara2024}, who evolved small separation, non-spinning black holes, i.e.,\ compact and weakly loaded mini-disks. At $30\,M$ separation, mini-disks are much more massive and do not completely deplete but are still sufficiently small to be strongly influenced by the sloshing gas.

\subsection{Mini-disk properties}

We now describe the time-averaged properties of the mini-disks measured in the frame of the BH as a function of radius, $\bar{r}$. Integrating over the last $10\,P_{\rm bin}$, we obtain inflow equilibrium up to a radius of $\bar{r} \simeq 6\,M$ {around} both black holes, with constant net mass flux $\langle \dot{M} \rangle \approx0.2 \,\dot{M}_0$. The measured surface-density profile{s} decay as $\Sigma\propto 1/\bar{r}^{2}$, substantially steeper than the $\Sigma\propto 1/\sqrt{\bar{r}}$ scaling expected for a steady thin
$\alpha$-disk with constant scale-height. This supports the fact that the mini-disks are not locally-steady viscous disks but are instead draining much faster {than the viscous timescale}. As we discussed in previous sections, various effect{s} increase the effective radial drift speed and reduce the amount of mass required to carry a given accretion rate.

{Our} entropy-target cooling prescription is calibrated to maintain the initial thickness of the circumbinary disk{,} $h/r\simeq 0.1$. The mini-disks, however, are not present in the initial data; they form self-consistently from streams entering an initially evacuated cavity, and therefore need not relax to the same constant scale height as the circumbinary disk. We find that the density-weighted scale height measured in coordinates centered on each black hole is $\langle \bar{h}/\bar{r} \rangle\approx 0.15$ when averaged over the mini-disk, increasing from 0.1 near the ISCO to 0.2 near the outer edge. This radial increase is consistent with stronger shock heating in the stream-impact and circularization regions: incoming streams deposit orbital energy in the outer mini-disk, and this heating occurs on timescales comparable to, or shorter than, the local (Keplerian) cooling time ($\sim 2\mathrm{\pi}/\Omega_{K}$). As a result, the gas cannot maintain the same entropy and thickness as the colder, turbulence-heated circumbinary disk.

\begin{figure}
        \centering
        \includegraphics[width=1.0\linewidth]{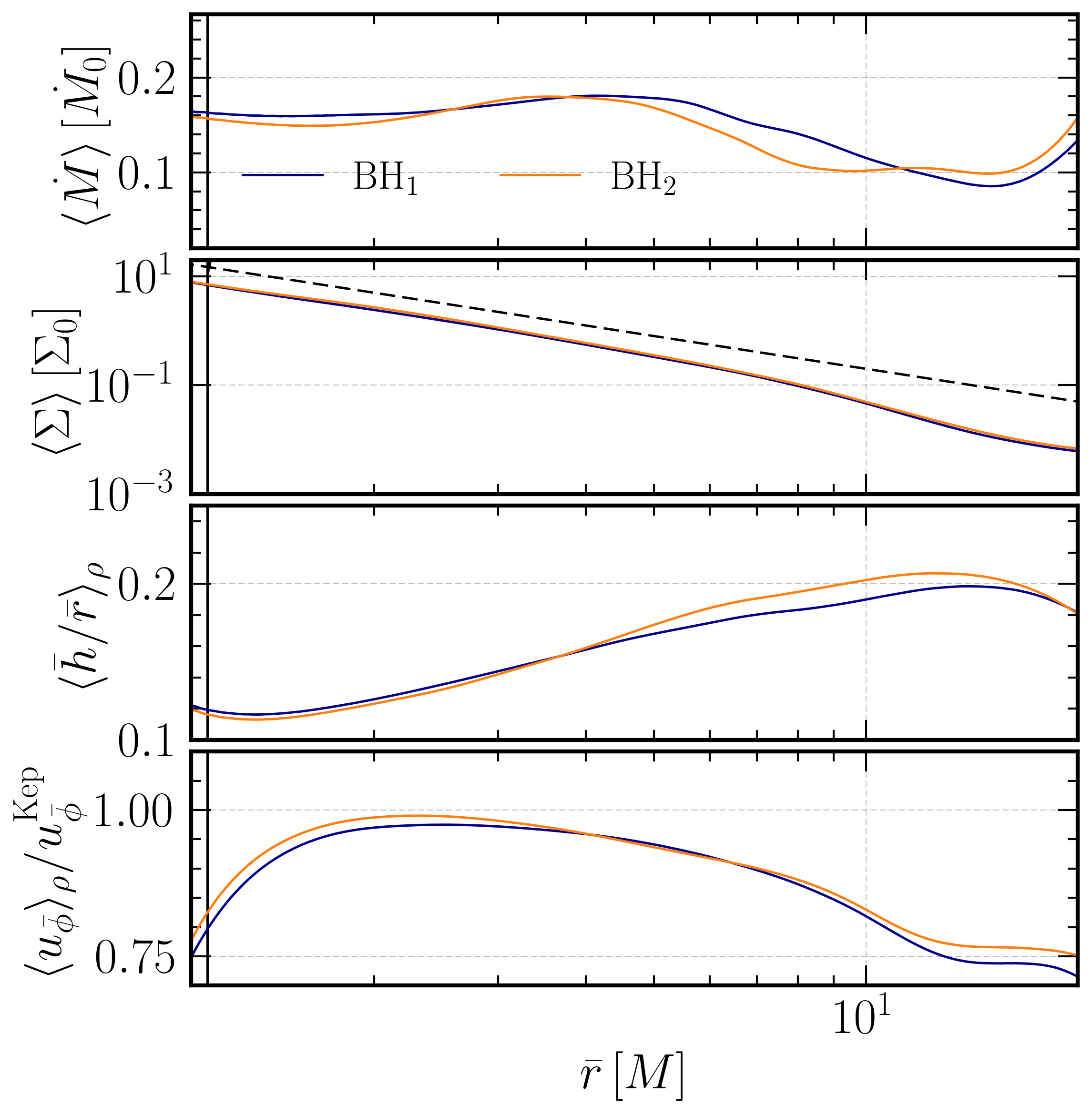}
        \caption{Time-averaged properties of {the} mini-disk{s} as a function of radius in the black hole frame; we show accretion rate (top panel), surface density normalized with $\Sigma_0=10^{-5}$ and dashed-lines showing $\propto r^{-2}$ (second panel), scale-height (third panel) and specific angular momentum, normalized by its Kerr circular-orbit value (bottom panel).}
        \label{fig:md_prop}
\end{figure}

Gas supplied by the circumbinary disk enters each mini-disk with a distribution of specific angular momentum that is sufficient{ly} low to overcome {the} centrifugal barrier of the binary. We can define an effective circularization radius $\bar{r}_{\rm circ}\sim j^2/m_{A}$, where $j$ is the specific angular momentum in the BH$_{A}$ frame of the incoming stream. {A crude estimate of $j$ follows by treating the stream as ballistic after leaving the cavity edge. Using the characteristic infalling stream angular momentum in the center of mass frame, $j_{\rm CM}\simeq1.4\sqrt{Mr_{12}}$, inferred from circumbinary-flow simulations \citep{Shi2015, Tiede2022} and transforming to the BH frame, we obtain $j\simeq 0.3\sqrt{Mr_{12}}$ and thus $\bar{r}_{\rm circ} \simeq 5.5\,M$ for $r_{12}=30\,M$, comparable to the outer edge of the region that has reached inflow equilibrium over the averaging interval}. Because the incoming flow shocks and joins the mini-disk at larger radii than $r_{\rm circ}$, the density-weighted angular-momentum distribution is naturally sub-Keplerian (Fig.~\ref{fig:md_prop}, bottom panel). Dissipation and angular-momentum redistribution progressively circularize the flow inward, and for $\bar r\lesssim5,M$ the mini-disk approaches the circular-orbit profile of an isolated Kerr black hole down to the ISCO {at $1.4\,M$}.

\subsection{Magnetic transport: from the CBD to the mini-disks}

\begin{figure}
    \centering
    \includegraphics[width=1.0\linewidth]{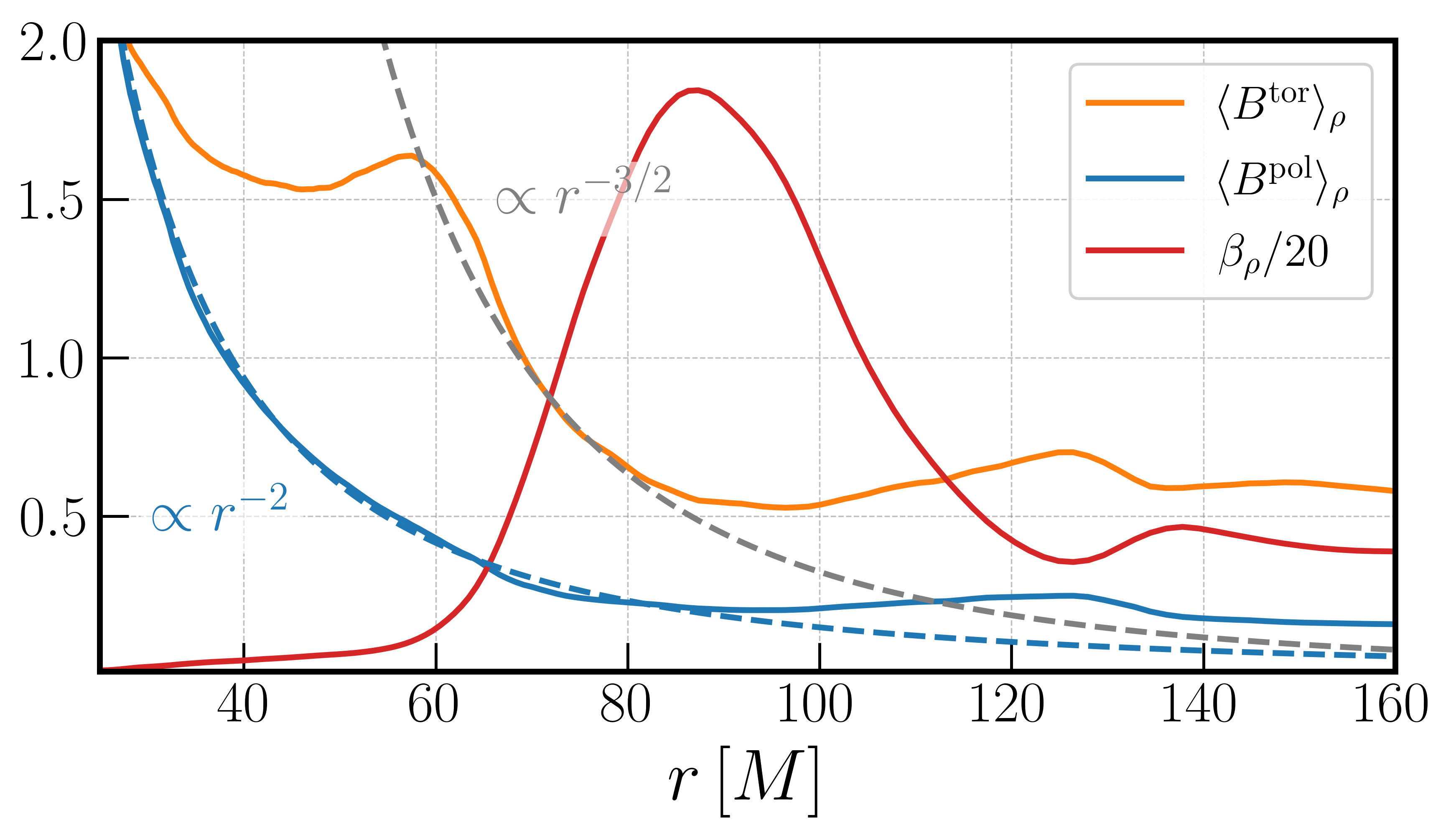}
    \caption{Spherically-averaged over time and density weighted plasma$-\beta$, defined as $\beta_{\rho}=2 \langle p\rangle_{\rho}/\langle b^2 \rangle_{\rho}$, poloidal ($B^{\rm pol}$) and toroidal $(B^{\rm tor})$ magnetic fields, as a function of radius.}
    \label{fig:bth_bph_phi_cbd}
\end{figure}

Ordered large-scale poloidal field near a BH engine can extract its rotational energy and power relativistic jets \citep{blandford1977Electromagnetic, komissarov2009BlandfordZnajek}. The transport and generation of magnetic fields from large radii to the horizon depend on properties of the accretion disk such as its scale height, angular momentum, and initial magnetization \citep{beckwith2008Influence,Beckwith2009, Hogg2018, Ressler2021, Galishnikova2025}.  The distinctive morphology of a binary accretion disk is then expected to profoundly change how the field is advected and amplified, especially for equal-mass binaries where torques are strongest. 

The turbulent transport of magnetic flux operating in single-BH disks cannot operate efficiently across the CBD inner edge. Instead, flux freezing advects the field into the cavity through accretion streams that stretch and compress it. The field is subsequently incorporated into the mini-disks or redistributed into the low-density cavity by shocks and outflows, allowing the cavity to become strongly magnetized. Because the CBD in our simulations carries little net magnetic flux, neither the cavity nor the black-hole horizons reaches a saturated, magnetically arrested state \citep{Most2024,Manikantan2025}. We analyze below the magnetic-field structure and transport in this standard accretion regime.

\begin{figure*}[ht!]
        \centering
        \includegraphics[width=.48\linewidth]{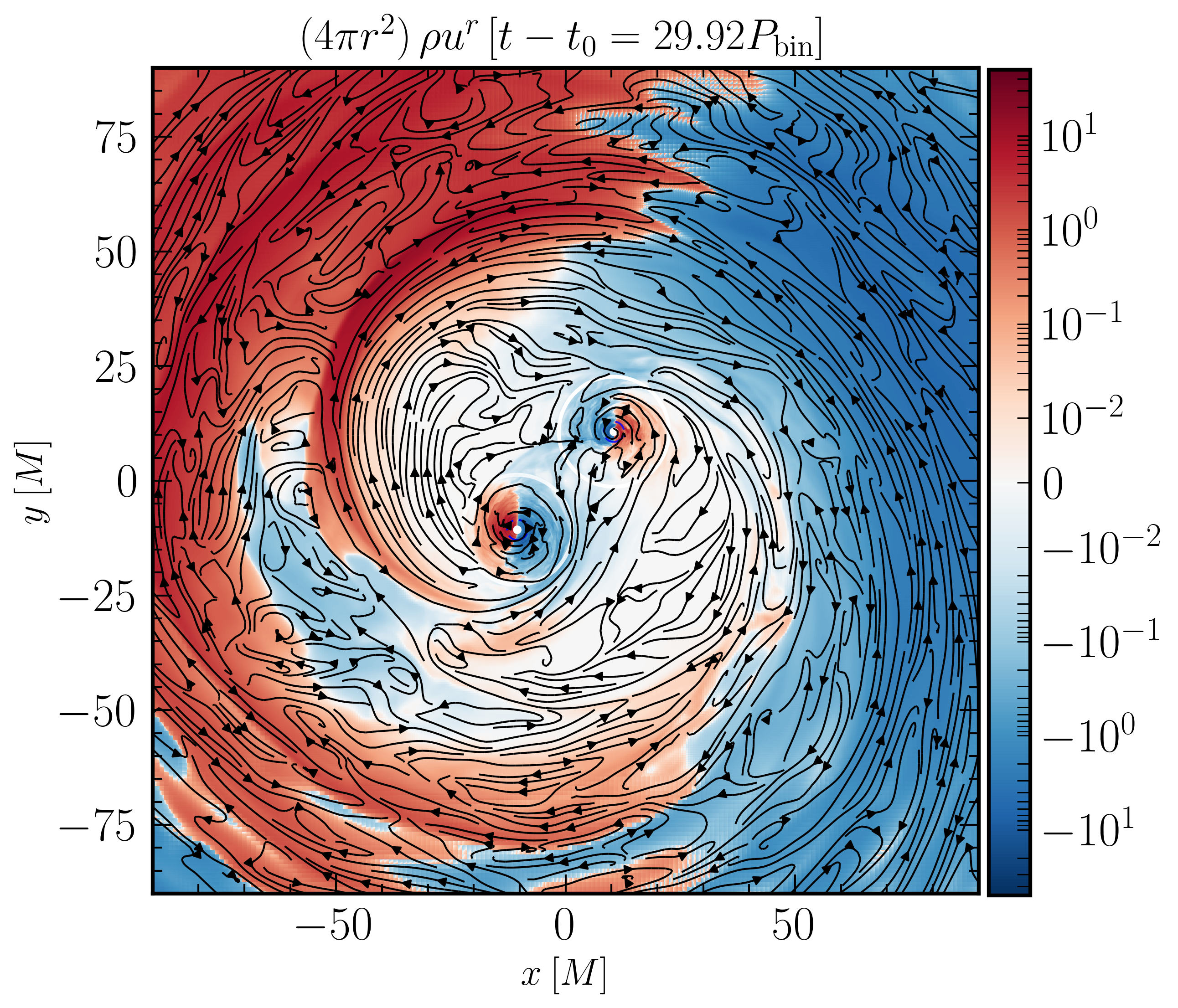}
        \includegraphics[width=.48\linewidth]{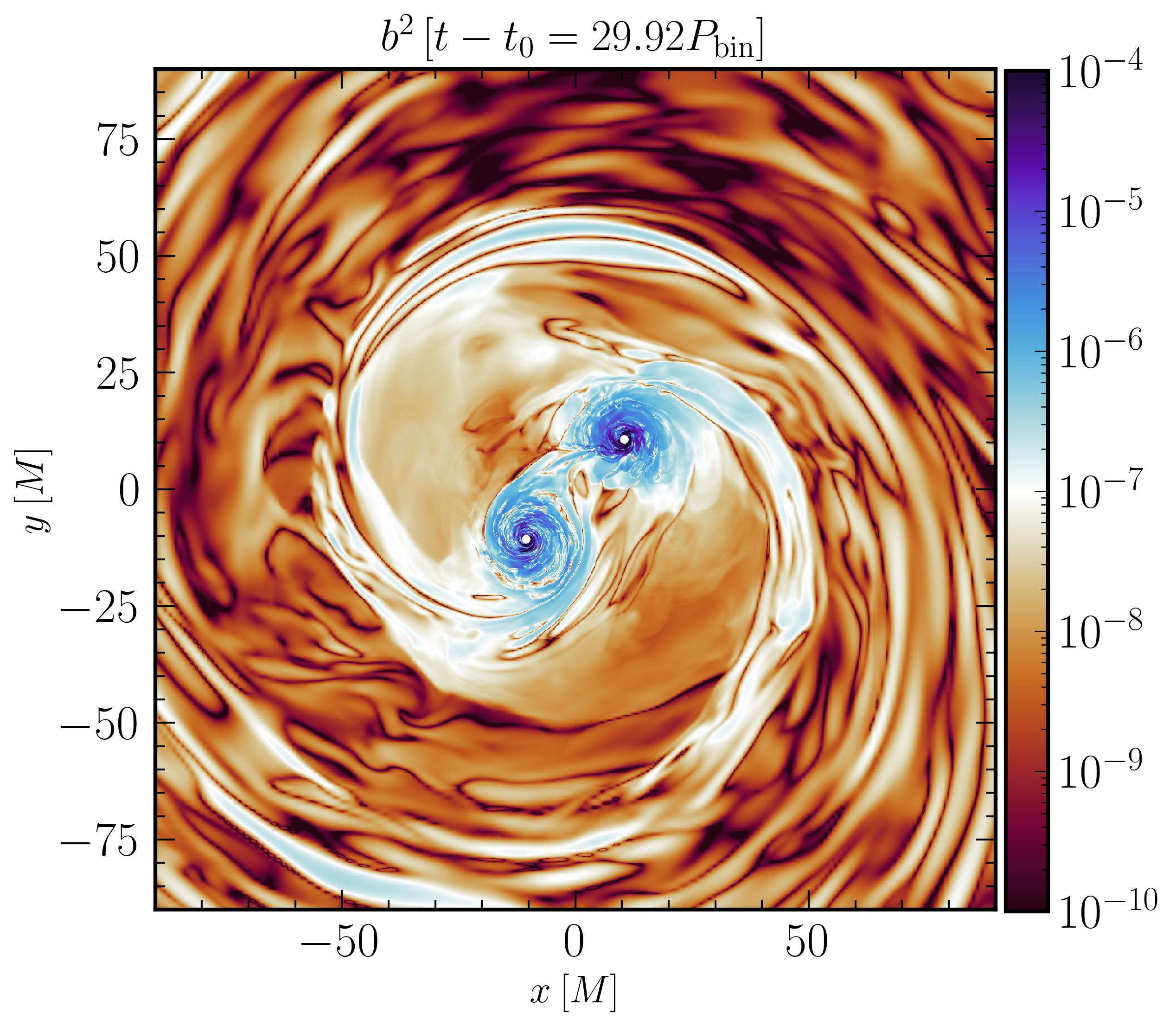}
        \includegraphics[width=.48\linewidth]{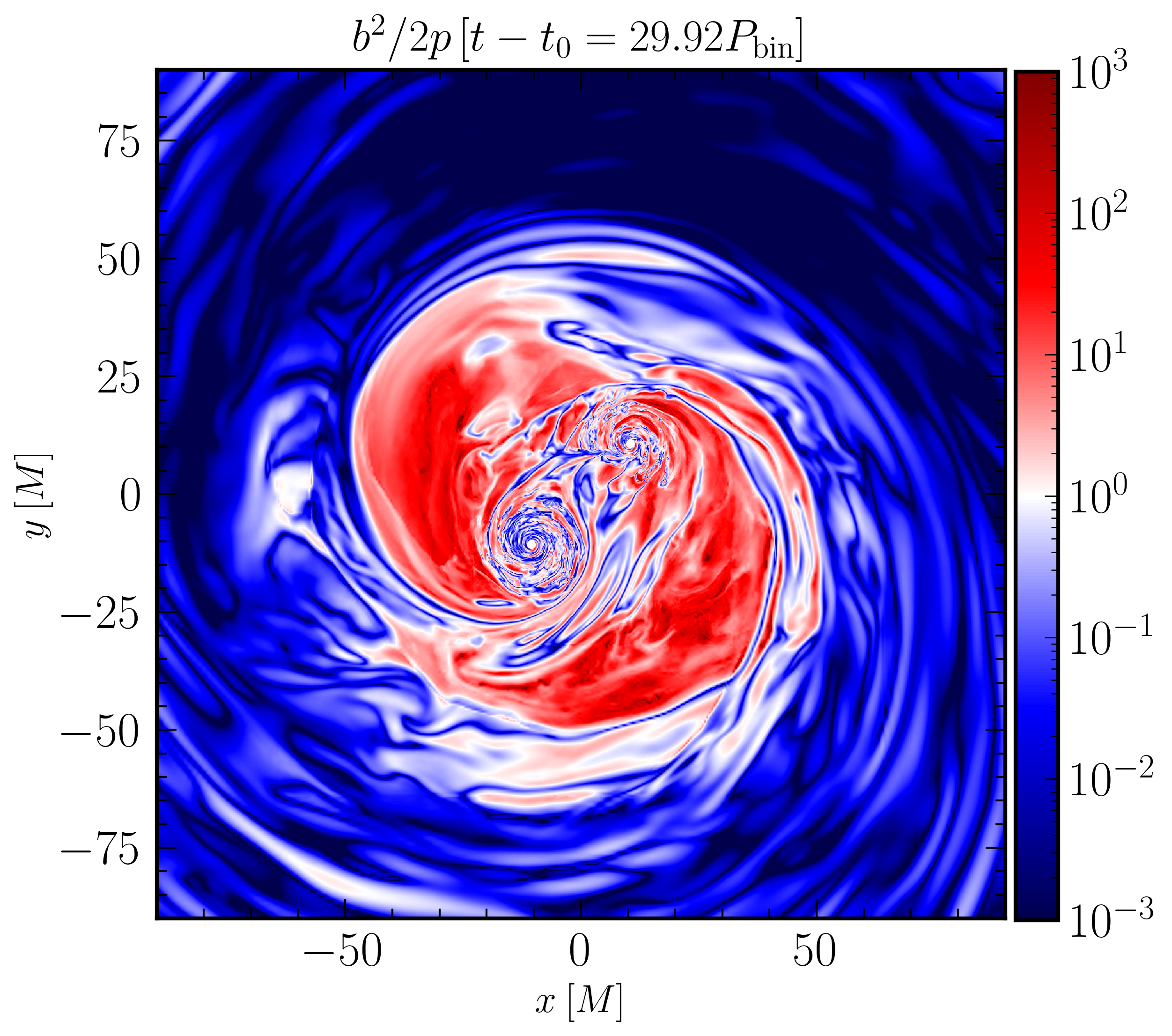}
        \includegraphics[width=.48\linewidth]{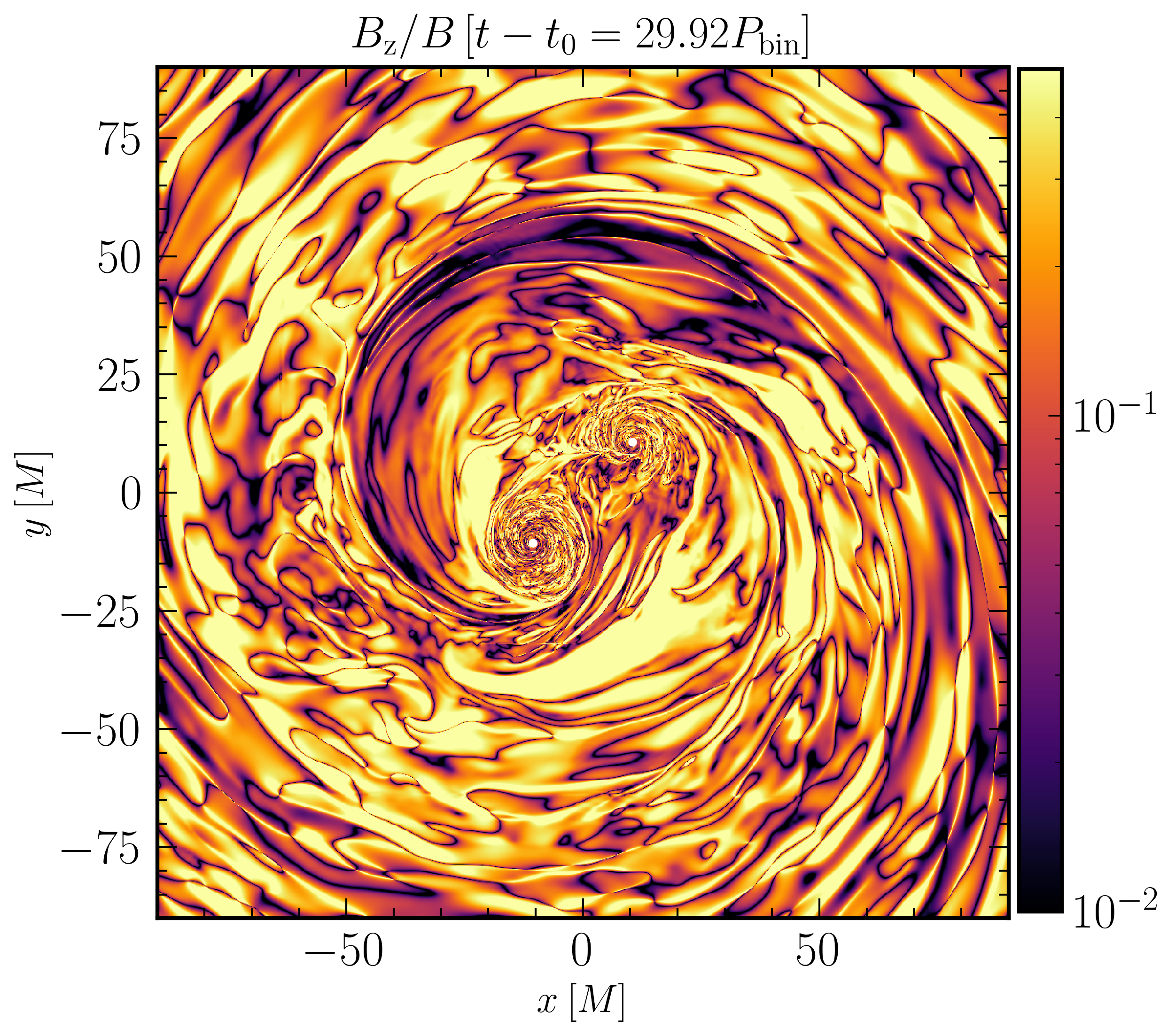}
        \caption{Equatorial snapshot showing mass flux with magnetic field lines (top left), comoving magnetic field energy density (top right), inverse plasma-$\beta$ (bottom left), and normalized vertical field (bottom right).}
        \label{fig:magfield_xy}
    \end{figure*}

We show time- and surface-averaged magnetic-field diagnostics as a function of radius in Fig.~\ref{fig:bth_bph_phi_cbd}. After reaching a quasi-steady state, the bulk of the circumbinary disk saturates at a plasma beta $\beta \equiv 2p/b^2 \sim 10$. The magnetic field is predominantly toroidal, with $\langle B^{\rm tor}\rangle_{\rho}/\langle B^{\rm pol} \rangle_{\rho}\sim 3$, broadly consistent with field topology found in MRI-driven disks \citep{hawley1995local, salvesen2016accretion}. Near the inner edge of the CBD, the magnetization decreases rapidly, reaching $\beta \sim 100$ at $r\sim 90\,M$, close to the overdense lump. 

Interior to $r\approx 90\,M$, the magnetic-field transport differs qualitatively from that in the turbulent CBD. The density-weighted poloidal field approximately follows $B_{\rm pol}\sim |B^r| \propto r^{-2}$, consistent with approximate conservation of radial magnetic flux within the stream-dominated flow. By contrast, the density-weighted toroidal field grows more slowly, approximately as $B_{\rm tor}\propto r^{-3/2}$, down to $r\approx 60\,M$, and then remains nearly flat until the flow reaches the mini-disks at $r\approx 25\,M$. This behavior suggests that the field in the cavity is transported primarily by coherent, laminar streams stripped from the CBD edge, rather than by turbulent flows.

The geometry and properties of the field in the equatorial plane are illustrated in Fig.~\ref{fig:magfield_xy}. When gas is stripped from the inner edge of the CBD, it initially carries too much angular momentum to be accreted directly by the binary. A fraction of this material is therefore torqued and flung back toward the CBD, visible as outgoing mass flux in the red contours of Fig.~\ref{fig:magfield_xy} (top left). In these outgoing streams, field lines are compressed into narrow regions and amplified, as shown by the enhanced comoving magnetic energy density in Fig.~\ref{fig:magfield_xy} (top right). The equatorial field lines in Fig.~\ref{fig:magfield_xy} (top left) show that the field becomes ordered and predominantly toroidal in the stream region, with a small relative vertical component, $B_z/B<1$, as shown in Fig.~\ref{fig:magfield_xy} (bottom right). The field is maximally compressed when the stream impacts the edge of the cavity, showing polarity inversion suggestive of reconnective dissipation. \cite{Noble2021} argued that magnetic dissipation in the lump region can weaken Maxwell stresses, allowing gas to accumulate without being efficiently sheared apart.

Once the gas loses enough angular momentum, it is ultimately accreted onto the binary through ingoing streams, shown by the blue contours in Fig.~\ref{fig:magfield_xy} (top left). In contrast to the outgoing recycled streams, the field in the accreting streams becomes more disordered as it approaches the mini-disks, and the toroidal and poloidal components become comparable, $B_{\rm pol}\sim B_{\rm tor}$, near the mini-disk edge (Fig.~\ref{fig:bth_bph_phi_cbd}). The low density cavity is strongly magnetized, $2p/b^2<10^{-3}$, as shown in  Fig.~\ref{fig:magfield_xy} (bottom left), but it is far from reaching a magnetically arrested state {where a significant amount of} large-scale coherent flux {gets} trapped in the cavity. Although we observe some patches containing {vertical} flux tubes, e.g., in the lower right corner of Fig.~\ref{fig:magfield_xy} (bottom right) where $B_z/B\sim1$, the coherence is intermittent and eventually lost on orbital time-scales. 

\subsection{Magnetic flux in the mini-disks}

\begin{figure}
    \centering
    \includegraphics[width=1.0\linewidth]{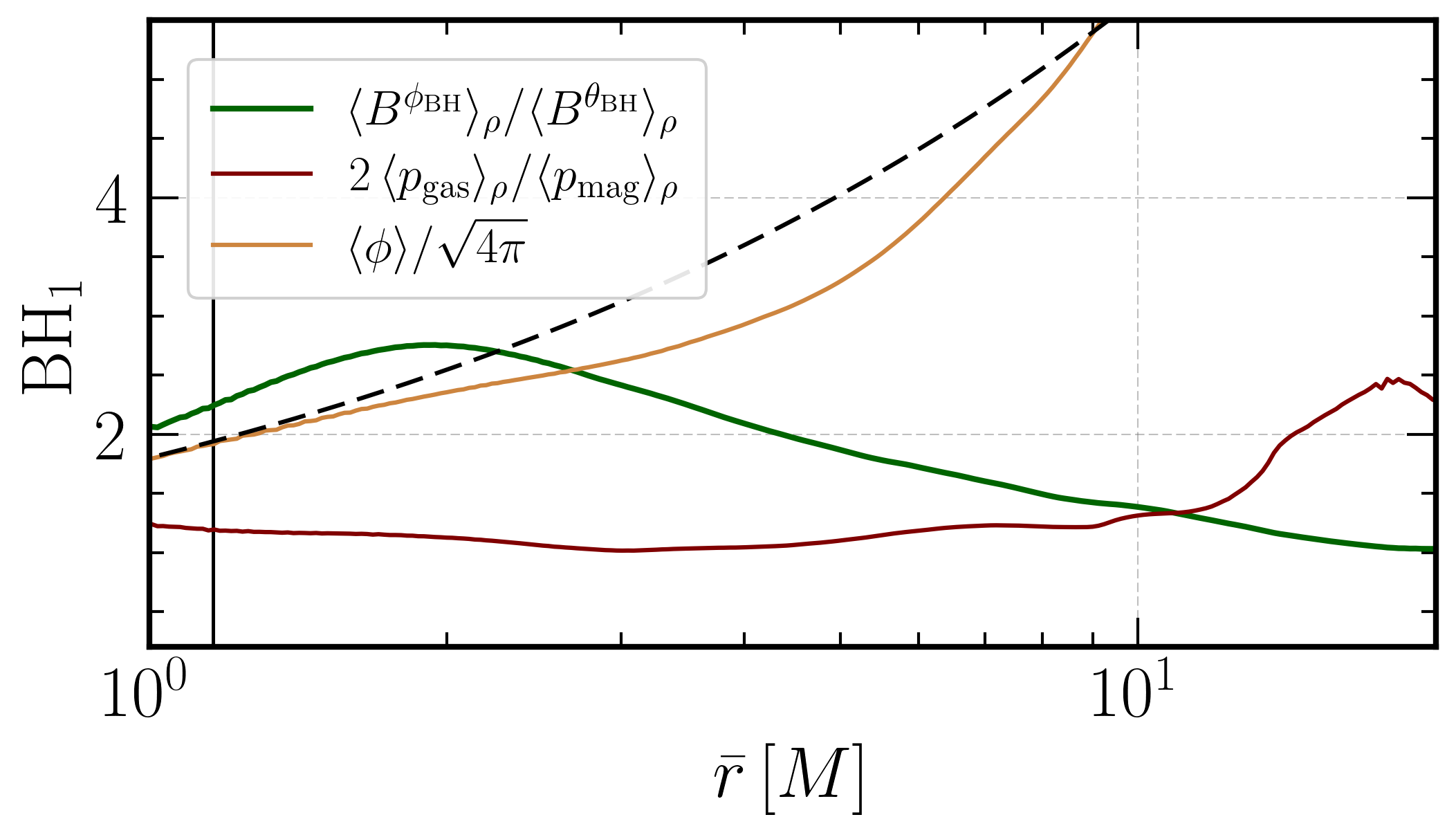}
    \caption{Dimensionless flux ($\phi$), poloidal ($B^{\theta'}$) and toroidal $(B^{\phi'})$ magnetic fields {vs.\ radius} in the frame of BH 1, spherically-averaged over time and density weighted.}
    \label{fig:bth_bph_phi_md}
\end{figure}

At the outer edge of the mini-disks, toroidal and poloidal field components in the BH frame are roughly equal (Fig.~\ref{fig:bth_bph_phi_md}), growing more toroidal near the black hole horizon with $B^{\rm tor}/B^{\rm pol}\approx 2$. The mini-disks are fairly magnetized, characterized by a time-averaged, radially constant plasma$-\beta$ of $\beta\approx1$, which is smaller than typical MRI-driven disks (see also $b^2/2p$ in Fig.~\ref{fig:magfield_xy}). Because mini-disks are lighter, thin, and more intermittent than a continuously fed standard disk, it is natural to expect a higher magnetization state. 

To measure the magnetic flux threading the horizons, we compute the unsigned
flux in the black-hole frame,
\begin{equation}
    \Phi_{\rm BH}
    =
    \frac{1}{2}\int_{r_{\rm H}} \sqrt{4\pi}\,
    |B^{\bar r}|\,d\bar A ,
\end{equation}
which we normalize with $\Phi_0=\sqrt{\dot{M}_0}$. We define the corresponding dimensionless magnetic flux as
$\phi_{\rm BH}=\Phi_{\rm BH}/\sqrt{\dot{M}_{\rm BH}}$.\footnote{The factor
of $1/2$ accounts for the integration over both hemispheres. The factor
$\sqrt{4\pi}$ converts our Eulerian magnetic field from Heaviside--Lorentz
units to the Gaussian-unit convention commonly used in definitions of
$\phi_{\rm BH}$.} We define the corresponding time-averaged value as
$\langle\phi\rangle(\bar{r})=\langle\Phi\rangle/
\sqrt{\langle\dot{M}\rangle}$. Near the horizon we find
$\langle\phi\rangle\simeq 7-15$, increasing outward approximately as
$\langle\phi\rangle\propto \bar{r}^{1/2}$
(Fig.~\ref{fig:bth_bph_phi_md}). This value is well below the magnetically-arrested disk (MAD) threshold,
$\phi_{\rm BH}\sim 40-60$, found in single-black-hole accretion, where the
accumulated poloidal flux becomes strong enough to impede
inflows and drive flux-eruption events
\citep{Igumenshchev2003,Igumenshchev2008,Narayan2003,
tchekhovskoy2011Efficient,McKinney2012,Begelman2022,Chatterjee2022}. The sub-MAD flux
level is expected from the modest poloidal flux in our initial
circumbinary disk. Although the mini-disks are substantially
magnetized, we do not observe flux eruptions of the type
seen in MAD simulations \citep{Ripperda2022, ressler2025, Manikantan2025}.

\begin{figure}
    \centering
    \includegraphics[width=1\linewidth]{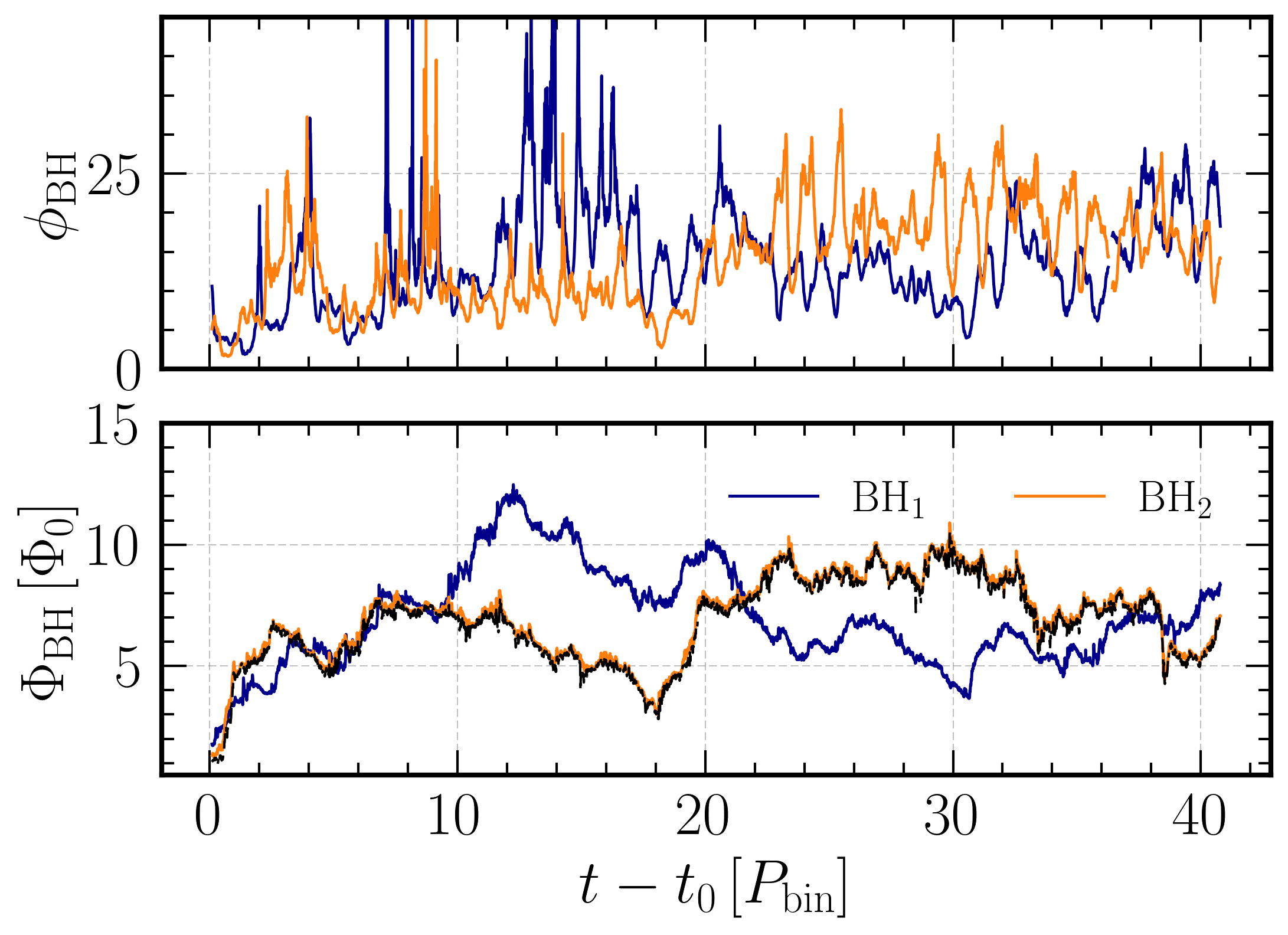}
    \caption{Top panel: dimensionless magnetic flux $\phi_{\rm BH}$ on each black hole as a function of time. Bottom panel: total magnetic flux onto each BH normalized with $\Phi_0$. The dominant magnetic flux alternates between each black hole. }
    \label{fig:mag_fluxes}
\end{figure}

We show the time evolution of the dimensionless and total unsigned flux in Fig.~\ref{fig:mag_fluxes}. During the first $10$ orbits, the absolute flux on the horizon grows similarly onto each black hole until reaching $\Phi\approx 7 \,\Phi_0$. After $10$ orbits, the flux starts decreasing on BH$_2$ but keeps growing on BH$_1$. This trend is reversed after $\approx 22$ orbits, when the fluxes on BH$_2$ starts dominating, and reverses once again after $\approx 36$ orbits. The alternating behavior is also present in the dimensionless flux $\phi$. We observe that $\phi$ exhibits orbital-scale periodicity, likely inherited from the accretion rate, with peaks of $\phi \approx 30$, especially clear for the dominating black hole{, i.e., the black hole with larger flux}.

The longer quasi-periodicity of the flux ($\sim 10$--$15\,P_{\rm orb}$) is not obviously associated with any of the hydrodynamical periods discussed above. Instead, it correlates most directly with the relative mass accumulated in the two mini-disks. The vertical flux delivered to a black hole is controlled by the competition between inward advection and outward diffusion of magnetic flux \citep{JacqueminIde2026,Ripperda2022}. In the present system, most of the mass and magnetic flux supplied by the lump is impulsively dumped into the cavity near the lump pericenter. Because the lump period is close to, but not exactly, an integer multiple of the binary period, the successive pericenter passages preferentially feed the same black hole for several lump cycles. The accumulated phase slip eventually changes which black hole is closest to the incoming stream, reversing the preferred feeding channel. This provides a natural explanation for the observed alternation of the horizon-threading flux between the two black holes on a timescale longer than both $P_{\rm orb}$ and the lump orbital period.  Note, however than in ideal MHD, diffusion and reconnection of field lines controlling the horizon-threading flux are dependent on resolution, and thus a full convergence study would be needed to establish how the flux alternates.

\begin{figure*}
        \centering
        \includegraphics[width=1.0\linewidth]{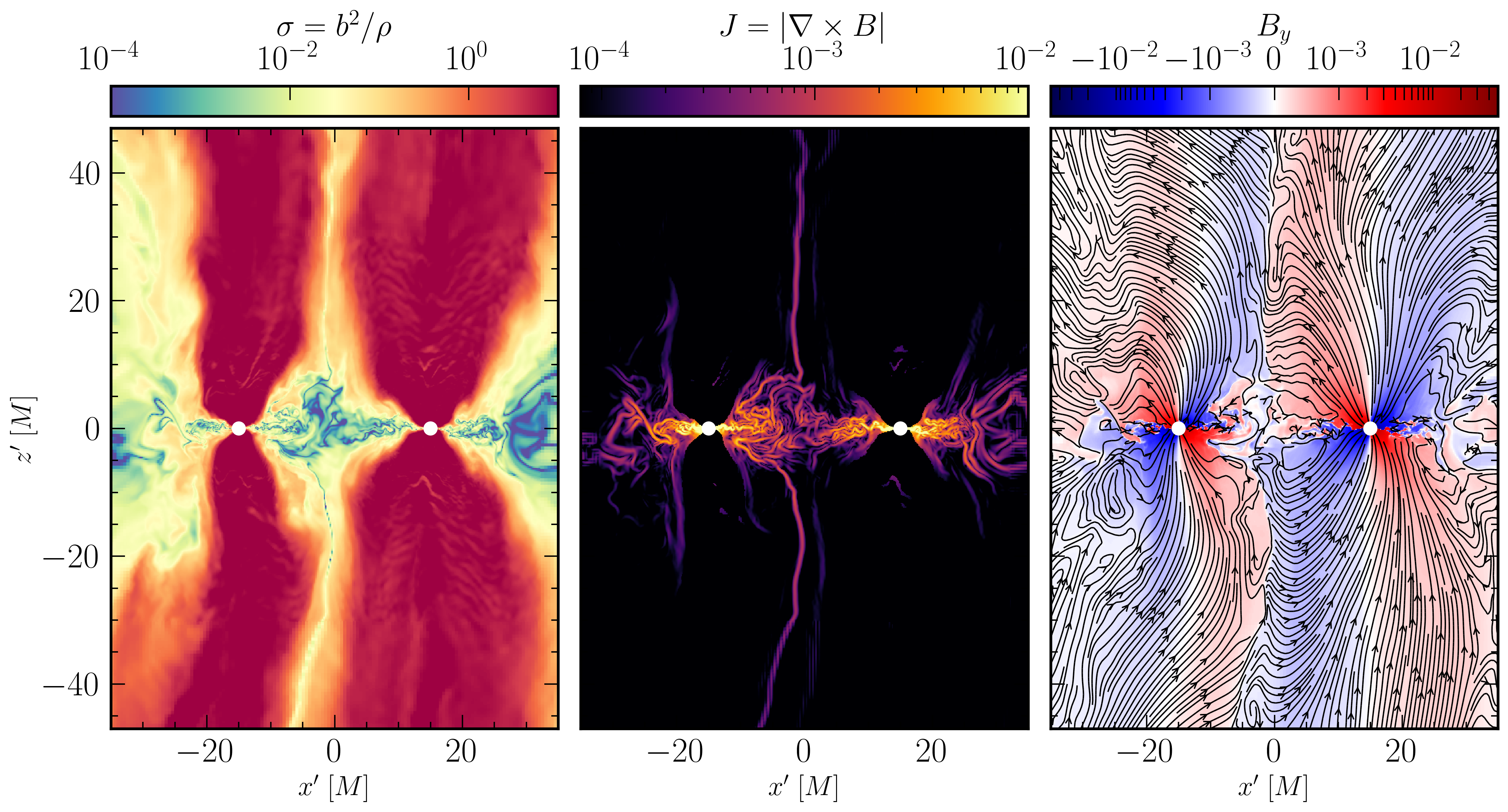}
        \caption{Structure of relativistic dual jets launched by each rapidly spinning BH in the corotating frame showing the magnetization (left), the norm of the ideal MHD current (center), and the toroidal field with field lines (right).}
        \label{fig:magfield_xz}
\end{figure*}

\subsection{Dual Poynting-dominated jets: morphology, luminosity, and interaction}

%Large-scale magnetic flux advected from the CBD anchors to the holes and launches dual Poynting-dominated jets powered by the BHs rotation, characterized by dimensionless spin $\chi =0.9$. The magnetic structure of the outflows is shown in the corotating plane of the binary in Fig.~\ref{fig:magfield_xz}. Ergospheric frame-dragging produces strong toroidal field near each black hole developing a steady dual magnetic tower structure, as observed in the field lines and perpendicular magnetic field component $B^{y'}$ in Fig.~\ref{fig:magfield_xz}A. The radial poloidal field lines threading the horizon bend quickly away from the center of mass axis, exhibiting a geometrical effect of the rotating system. 

%A highly magnetized ($\sigma=b^2/\rho \gg1$) evacuated funnel forms around each black hole, clearing out a very wide opening angle of ($\theta_{\rm jet} \gtrsim 50 \deg$) near the holes, which collimates into cylindrical shape further out at $z\gtrsim 15$. Because the mini-disks are thin and small, stronger collimation occurs further away, determined by the ram pressure of the outer CBD/cavity flows, companion jet, and outflow material from the sloshing region; see also Fig.~\ref{fig:rho_2d})B.

Large-scale magnetic flux advected inward from the circumbinary disk accumulates and threads the horizons of both black holes, launching two Poynting-dominated jetted outflows powered by the rapid black-hole rotation ($\chi=0.9$). Figure~\ref{fig:magfield_xz} shows the magnetic structure in the $x'$--$z'$ plane corotating with the binary.  Frame dragging twists the open poloidal flux and generates a strong out-of-plane component $B^{y'}$ near each black hole. In this plane, along each funnel, $B^{y'}$  acts as a proxy for the toroidal field wound around the jet, see also Fig.~\ref{fig:dualjets_3d} for a 3D view of the field lines. 

A highly magnetized, evacuated funnel with $\sigma=b^2/\rho\gg1$ forms around each black hole. Close to the horizons the funnel boundary subtends a large opening angle ($\theta_{\rm jet}\gtrsim 50^\circ$) because the mini-disks are thin and compact and therefore provide little hydrodynamic confinement. The outflows collimate only at larger heights ($z'\gtrsim 15M$) where the funnel region approaches a cylindrical morphology. This collimation is likely set by the pressure balance with the ram and thermal pressure of the circumbinary gas, the magnetic pressure of the companion jet, and material expelled from the sloshing region (Fig.~\ref{fig:rho_2d}, left).
\begin{figure*}
    \centering
    \includegraphics[width=1\linewidth]{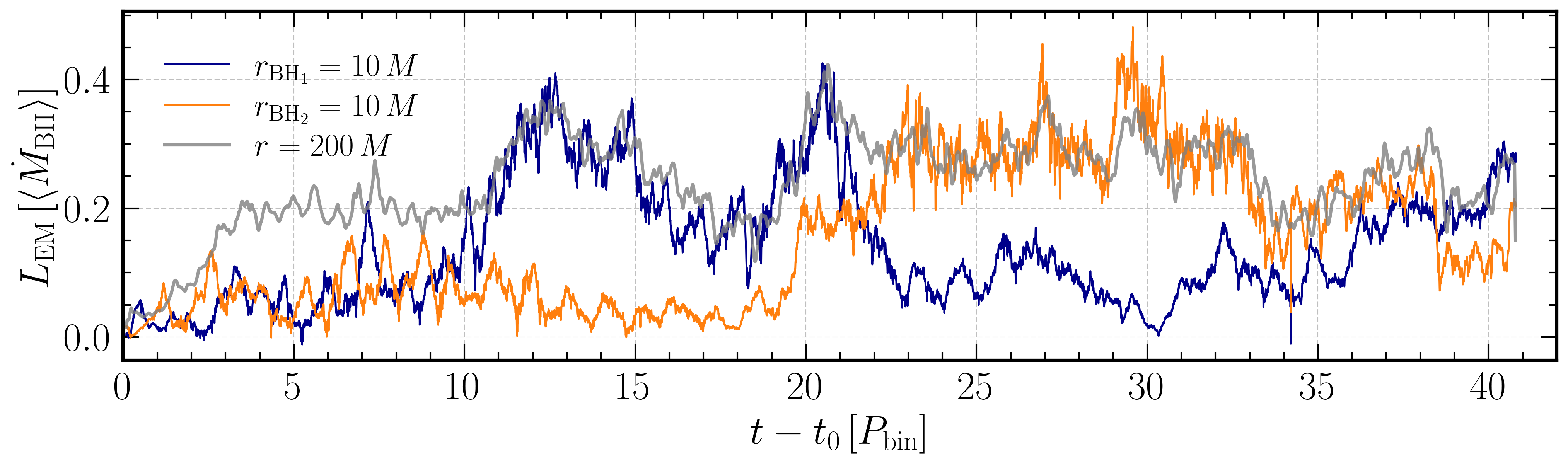 }
    \caption{Poynting luminosity normalized by the time-averaged accretion rate onto the black holes. Blue and orange curves show the near-zone electromagnetic power measured on spheres of radius $r_{\rm BH}=10M$ centered on BH$_1$ and BH$_2$, respectively. The gray curve shows the net luminosity measured on a center-of-mass sphere at $r=200M$. The far-zone luminosity remains roughly steady at $\eta_{\rm EM}\sim 0.2$--$0.3$, while the near-zone contribution alternates between the two black holes.}
    \label{fig:poynting}
\end{figure*}

\begin{figure}[ht]
        \centering
        \includegraphics[width=1.0\linewidth]{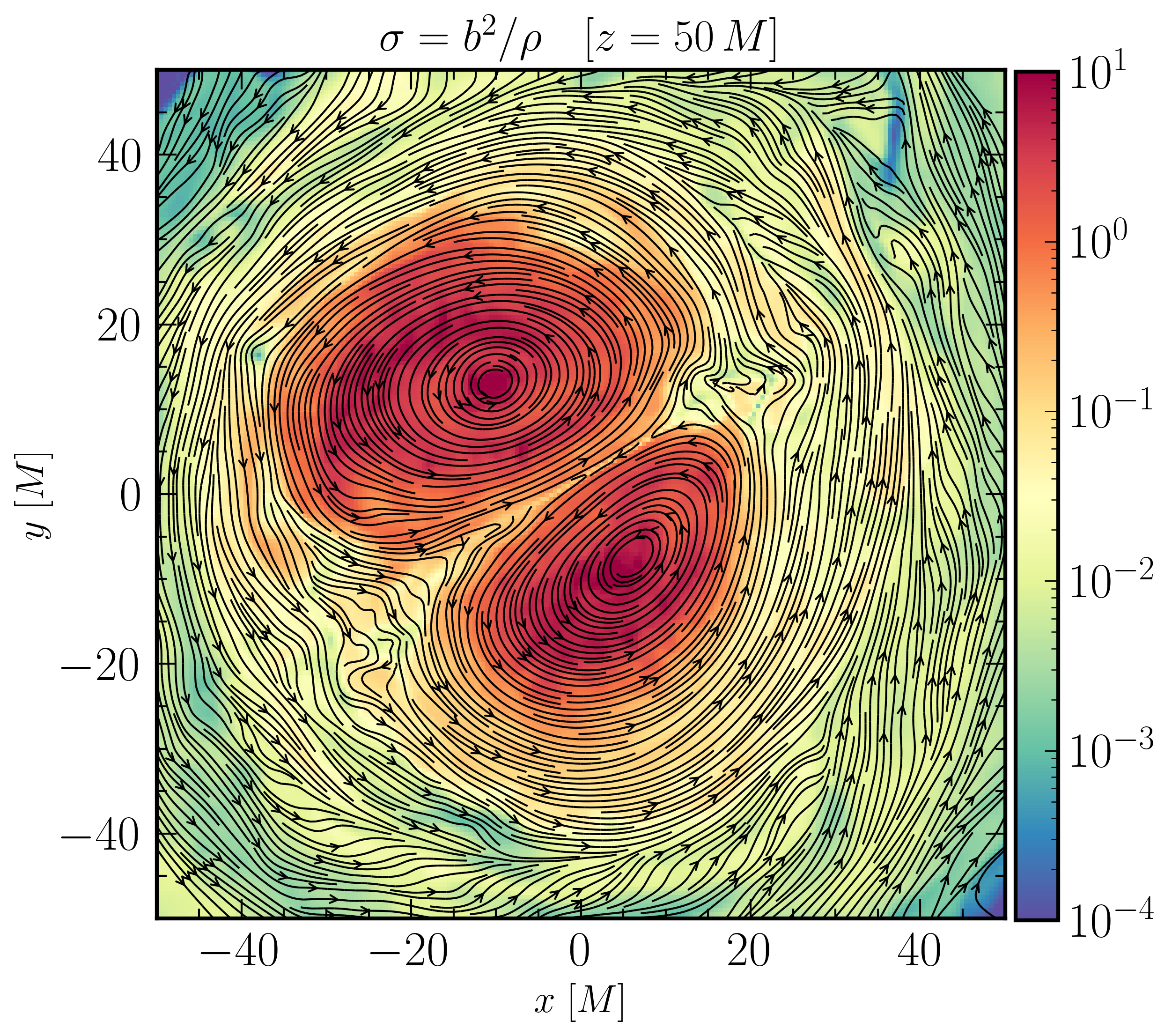}
        \caption{Magnetization and magnetic field lines over a slice at $z=50\,M$ above the binary. The dual jet structure appears as interacting flux tubes that reconnect in the middle region.}
        \label{fig:sigma_z50}
\end{figure}

\begin{figure*}[ht]
        \centering
        \includegraphics[width=0.48\linewidth]{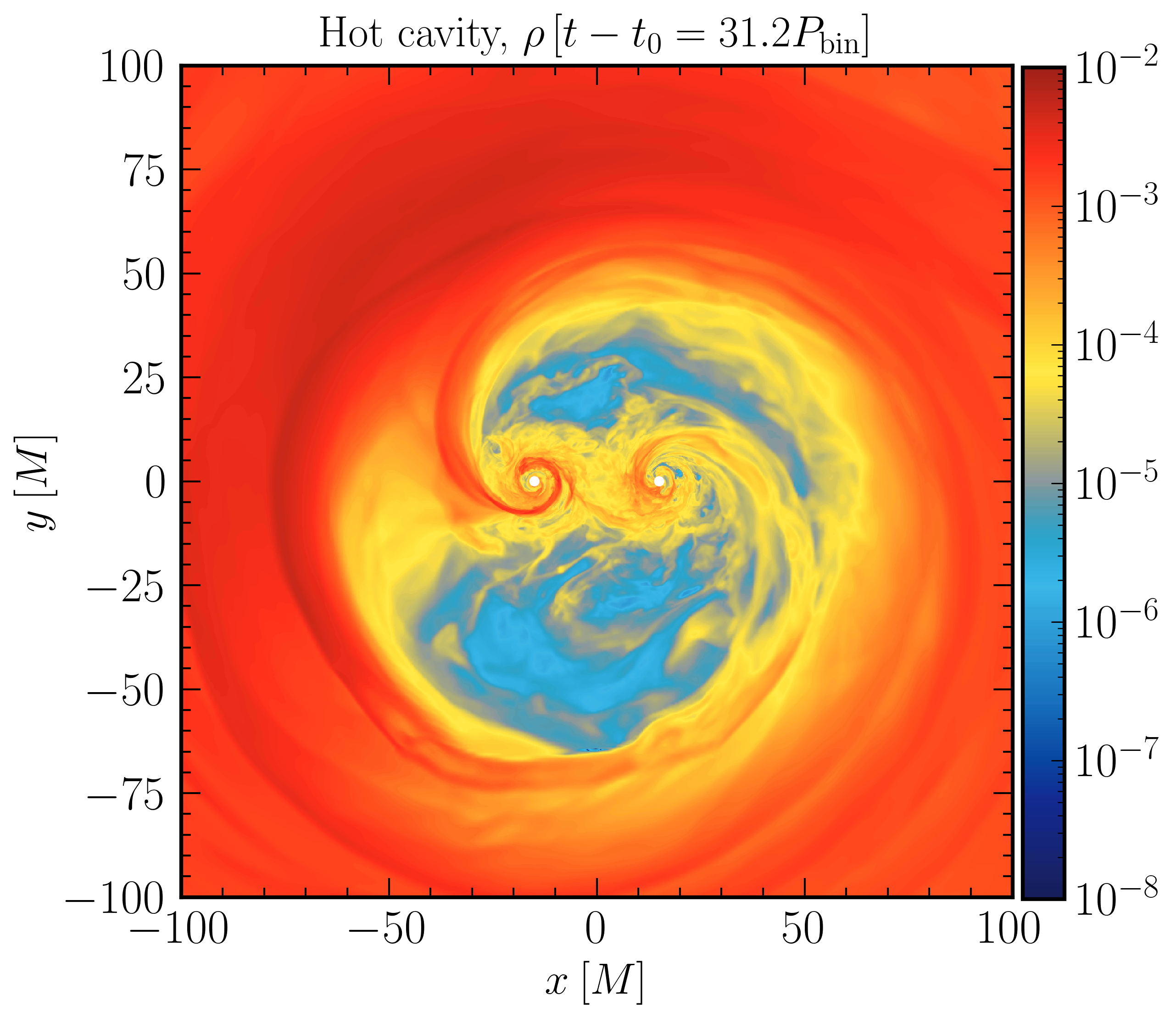}
        \includegraphics[width=0.48\linewidth]{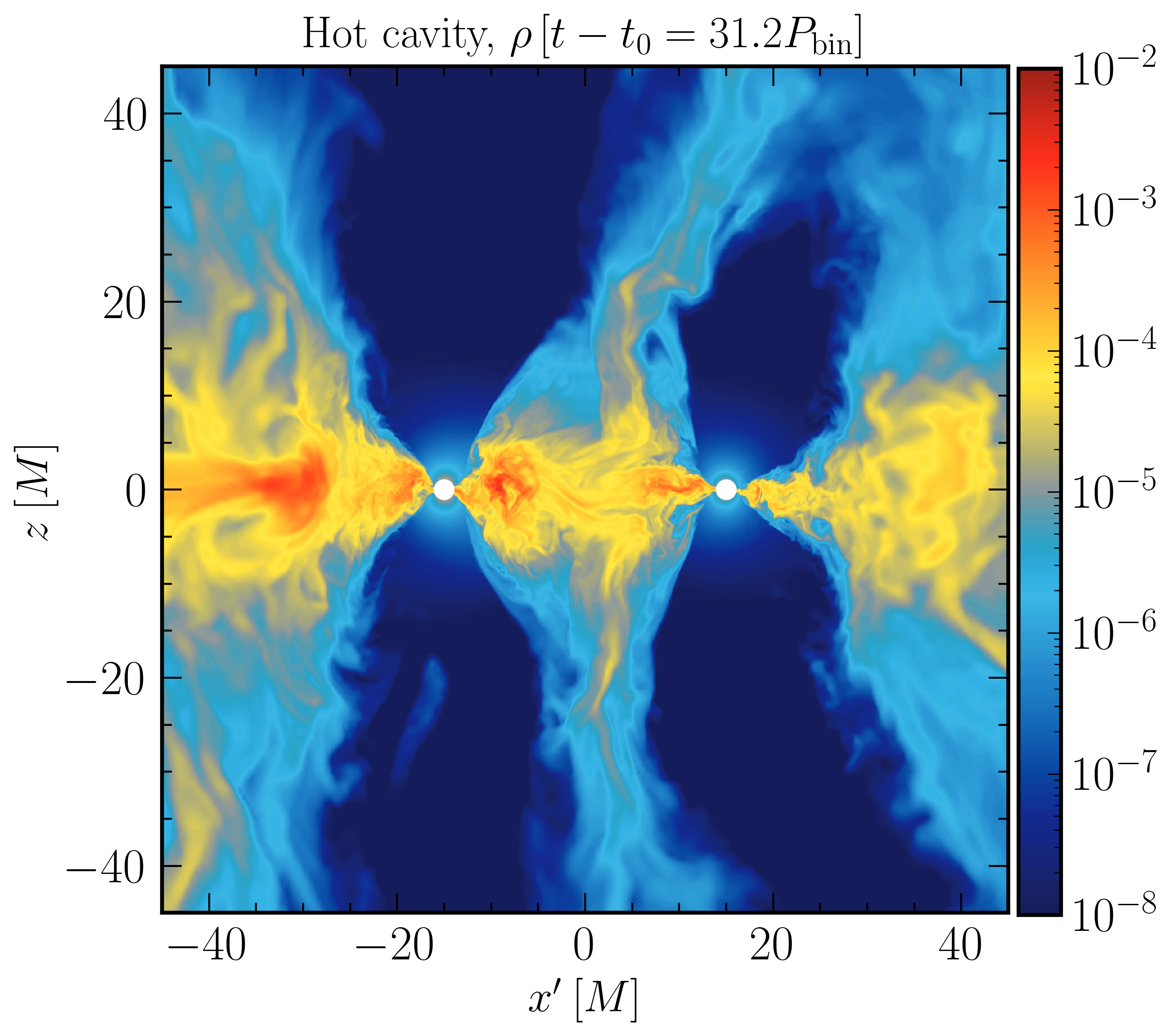}
        \caption{Equatorial (left) and meridional (right) snapshots of rest-mass density for the simulation with a hot cavity (without cooling the interior). }
        \label{fig:hot_rho_2d}
\end{figure*}

\begin{figure}
        \centering
        \includegraphics[width=1.0\linewidth]{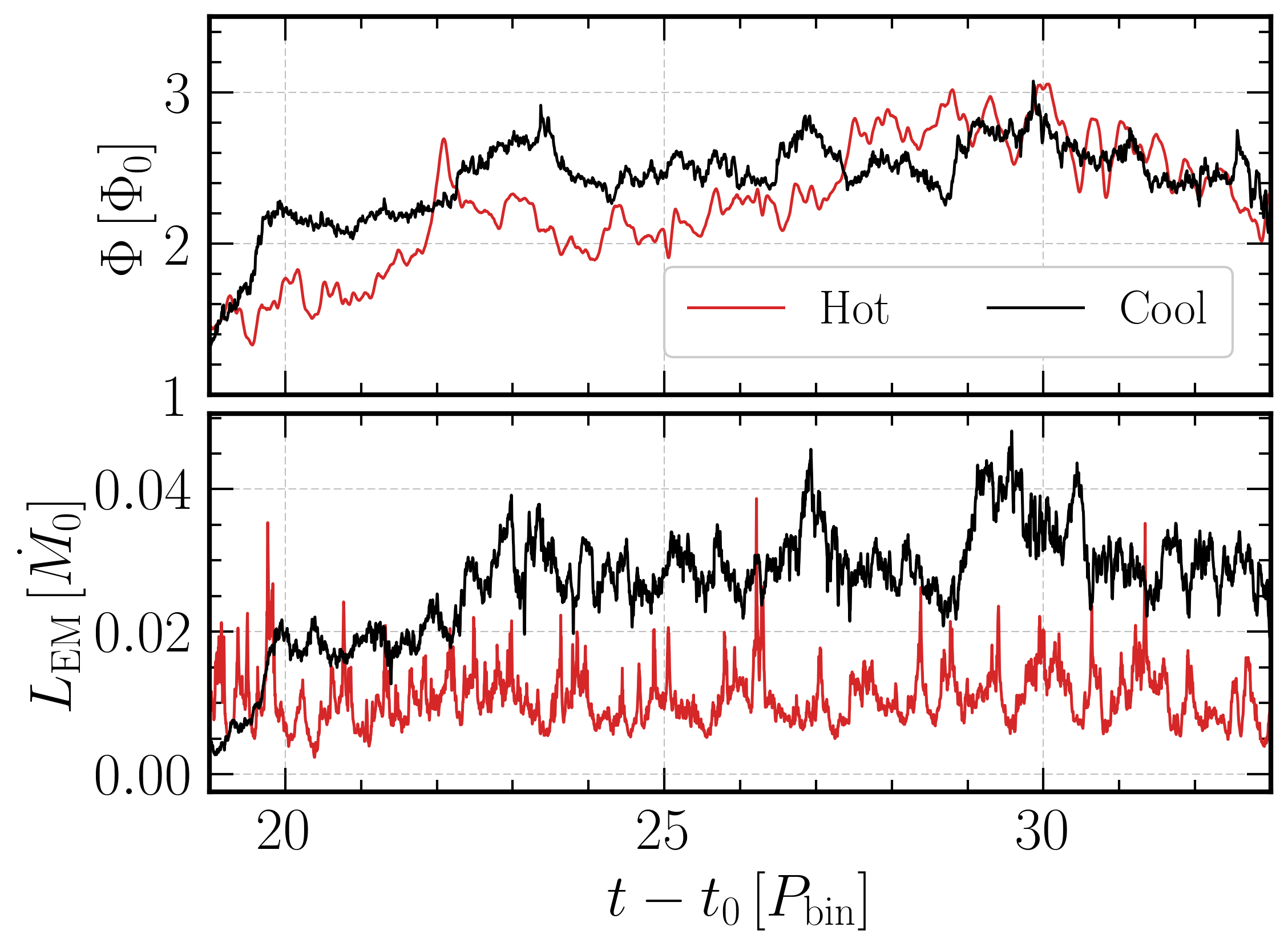}
        \caption{Comparison between simulations with hot and cold mini-disks for magnetic flux onto the horizon of BH$_2$ (upper panel) and Poynting flux luminosity (lower panel).{Curves for BH$_1$ show similar trends.} }
        \label{fig:hot_cold_bfield}
\end{figure}

\begin{figure}
        \centering
        \includegraphics[width=1.0\linewidth]{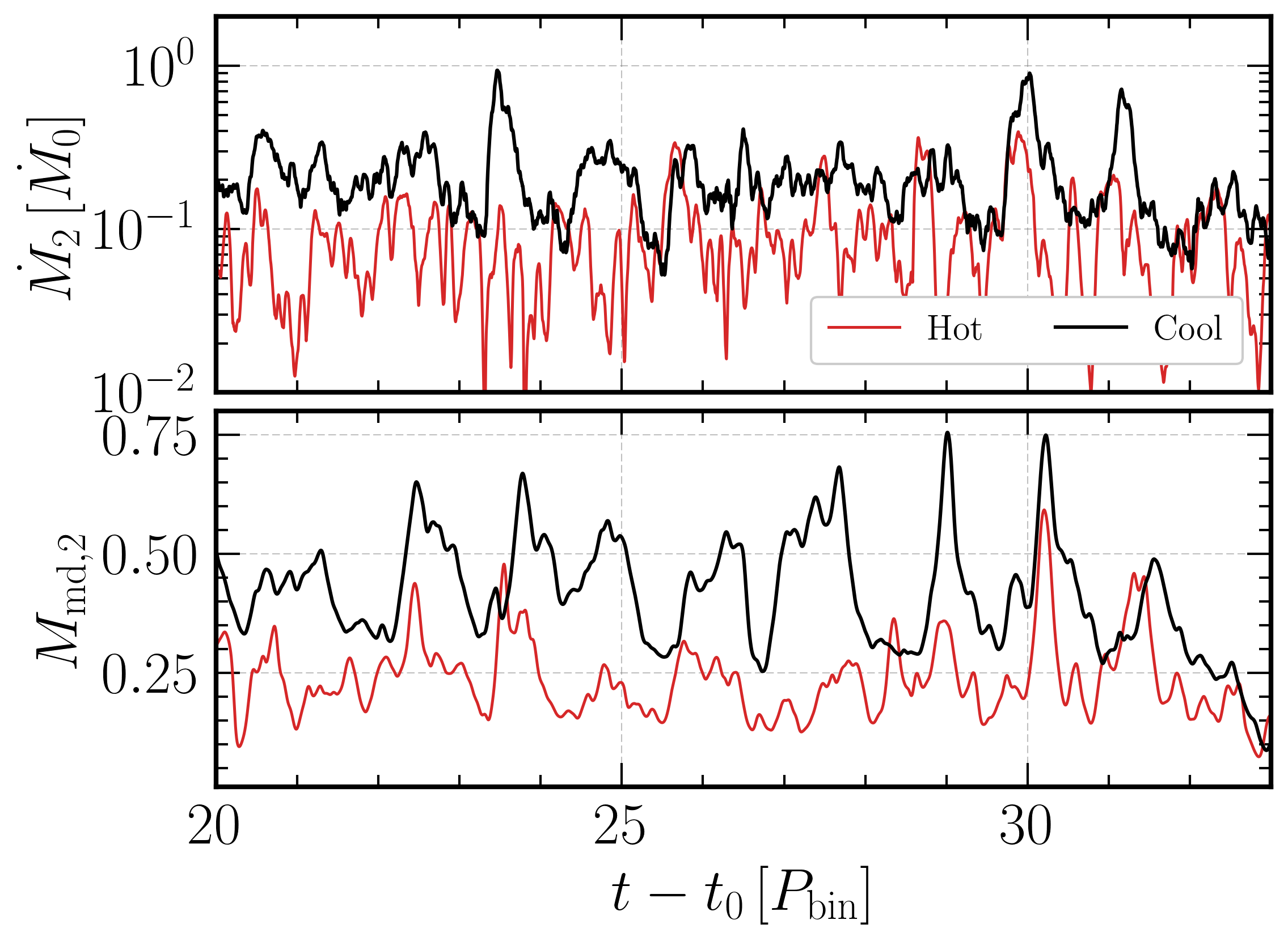}
        \caption{Comparison between simulations with hot and cold mini-disks for accretion rate onto the horizon of BH$_2$ (upper panel) and mass of mini-disk (lower panel). {Curves for BH$_1$ show similar trends.}}
        \label{fig:hot_cold_mass}
\end{figure}

The horizon-threading poloidal flux and the black-hole spin set the
electromagnetic power from these systems. We measure the electromagnetic
luminosity as the net outward energy flux,
\begin{equation}
    L_{\rm EM}
    =
    -\int \left(b^2 u^r u_t - b^r b_t\right)\,dA ,
\end{equation}
where we integrate the radial flux of the electromagnetic part of the stress-energy tensor. We compute this quantity on spherical surfaces centered on each black hole, at $\bar{r}=10\,M$, just outside the mini-disks, and on a larger sphere centered on the binary center of mass at $r=200\,M$. The former measures the near-zone electromagnetic power arising from each black hole, whereas the latter measures the luminosity that escapes to large radius. We normalize the luminosities by the time-averaged accretion rate onto the black holes, defining an effective electromagnetic outflow efficiency.

Figure~\ref{fig:poynting} shows that the large-radius luminosity rapidly reaches a quasi-steady value, $L_{\rm EM}\simeq 0.3\,\langle \dot{M}_{\rm BH}\rangle$, after the first few binary orbits. The net luminosity at large scales is roughly the sum of the near-zone contribution of each black hole, showing that most of the power arises from the BHs and can propagate out. The total electromagnetic power shows weak orbital variability, much less pronounced than simulations with depleting mini-disks \citep{Combi2022}. During the first $\sim 10$ orbits, both black holes contribute comparably to the net luminosity, with individual luminosities $L_{\rm EM}\lesssim 0.1\langle \dot{M}_{\rm BH}\rangle$. Around $t-t_0\simeq 10\,P_{\rm bin}$, the luminosity from BH$_1$ increases by a factor of several, reaching $L_{\rm EM}\simeq 0.3$--$0.4\,\langle\dot{M}_{\rm BH}\rangle$, while the contribution from BH$_2$ remains below $\sim 0.1$. The dominant source then switches to BH$_2$ near $t-t_0\simeq 22P_{\rm bin}$, when the BH$_1$ luminosity decays and BH$_2$ brightens to a comparable level. A further reversal to BH$_1$ occurs near $t-t_0\simeq 38P_{\rm bin}$. 

This alternating behavior closely follows the evolution of the magnetic flux threading each horizon, shown in Fig.~\ref{fig:mag_fluxes}. Since the spins are fixed and equal, the leading-order Blandford--Znajek scaling, $L_{\rm BZ}\propto \chi^2\Phi_{\rm BH}^2$, implies that variations in the jet power are primarily driven by changes in the horizon-threading magnetic flux. The dual outflows thus exhibit an ``on--off'' modulation in which one jet usually dominates the local electromagnetic output. The same alternation is visible in the instantaneous magnetic morphology.
In Fig.~\ref{fig:magfield_xz}, the dominating jet associated with the left black hole has a wider magnetized funnel and a more coherent vertical field structure, whereas the weaker jet is more strongly distorted by the inter-jet current sheet and the sloshing flow.  

Because the quasi-steady jets have large opening angles, the two funnels
interact continuously along the binary rotation axis for $z\gtrsim 15M$.
In this interaction region, the out-of-plane field component reverses sign across
the inter-jet layer. This toroidal field reversal is expected for aligned black-hole spins
threaded by magnetic flux of the same polarity. The
resulting sharp shear in the tangential magnetic field produces an extended
current sheet, visible as a strong
$J=|\nabla\times B|$ region in Fig.~\ref{fig:magfield_xz}.

The interface between the two funnels is therefore a natural site for
magnetic reconnection \citep{Gutierrez2024,ressler2025}. This is illustrated
by the transverse slice at $z=50M$ shown in Fig.~\ref{fig:sigma_z50}. The two
jets appear as interacting magnetized flux tubes, with cores of high
magnetization, $\sigma=b^2/\rho>1$, separated by a hotter, baryon-loaded
contact layer with $\sigma\sim 0.1$ as dense material from the sloshing region
enters this interface and lowers the local magnetization. The poloidal guide
field in this region is weak (Fig.~\ref{fig:magfield_xz}B), making the reconnecting component of the field close to anti-parallel. The projected in-plane magnetic field shows an
X-type topology at the interface, consistent with an extended reconnecting
current layer of characteristic width $\Delta x\approx 10M$.

The inter-jet current sheet  may
affect the large-scale jet morphology of the outflow by dissipating magnetic energy at {the} interface, and injecting baryon-loaded pressure fluctuations that perturb the collimated
outflows. The layer may also be susceptible to tearing or plasmoid formation,
although the development of such structures in our ideal-GRMHD calculation is
controlled by numerical resistivity and resolution{, and our resolution in this region is much lower than that needed to resolve individual plasmoids \citep{Ripperda2022}}. Higher-resolution
simulations, ideally with an explicit resistive or kinetic treatment, will be required to determine whether this reconnecting interface provides an observable probe of small-separation binaries.

\section{Simulations of cold and hot mini-disks accretion}

\subsection{Mini-disk thermodynamics}

The CBD and the gas inside the cavity need not lie in the same thermodynamical regime. A continuously-fed turbulent CBD can maintain a large surface density, remain optically thick, and cool efficiently. By contrast, the cavity gas and mini-disks are fed through narrow streams, can be shock-heated near circularization, and may have a short residence time because of their limited radial extent and efficient angular-momentum transport by tidal and spiral shocks. Mini-disks accreting from a radiatively efficient CBD will remain cold only if radiative losses can balance the local heating due to shocks and turbulence before the dissipated energy is advected into the black holes.

Following \citet{narayan1996New}, a hot, optically thin, advection-dominated equilibrium is possible below a critical accretion rate of order $\dot M_{\rm crit}\propto\alpha^2\dot M_{\rm Edd}$, where $\alpha$ is the viscosity parameter and $\dot{M}_{\rm edd}$ is the Eddington accretion rate. This model assumes a two-temperature, optically thin plasma in which ions receive most of the dissipative heating, electrons are heated through Coulomb coupling, and electron cooling includes synchrotron, bremsstrahlung, and Comptonization. For a finite-size mini-disk, the critical accretion rate scales as
\begin{equation}
\dot M_{\rm crit}
\sim
10^{-3}
\left(\frac{\alpha}{0.1}\right)^2 \,\Big(1-\sqrt{r_{\rm ISCO}/r_{\rm t}}\Big)^{-2}\,\dot{M}_{\rm Edd},
\end{equation}
where we have assumed a truncated steady $\alpha
$-disk and the limit where viscous heating equals the rate of energy transfer from the ions to the electrons. As the binary shrinks, the mini-disk inflow time decreases and the dissipation due to shocks can become larger than that of a typical viscously turbulent disk, potentially increasing the critical accretion rate by more than an order of magnitude. At sufficiently small separations, gas inside the cavity could therefore become radiatively inefficient even while the circumbinary disk remains cool and optically thick.  

We have assumed above that the accretion rate through the cavity  remains constant. If, on the other hand, the circumbinary disk decouples from the cavity due to the rapid GW-driven inspiral of the binary \citep{Milosavljevic2005,Armitage2002, Dittmann:2023dss,Most2025}, the accretion rate will drop, and the mini-disk may also naturally starve and become hot. Similarly, for unequal mass binaries, the primary black hole receives a smaller portion of the total mass flux, which could also drive the mini-disk to a hot state as explored by \cite{Tiede2025}. In the late inspiral regime, the thermodynamical state and radiative properties of the mini-disks can thus be very different than expected. 

In this section, we explore this mixed regime for the first time by comparing the physics of hot and cold mini-disks accreting from a cold circumbinary disk.

\subsection{Mass and electromagnetic fluxes of hot and cold mini-disks}

We now show the results of our simulation aimed at modeling a mixed thermodynamic regime in which the circumbinary  disk is kept efficiently cooled, while the gas in the mini-disks and inner cavity is allowed to retain heat. As we discussed above, the setup is motivated by the possibility that the lower-density mini-disks may not have sufficient optical depth, surface density, or cooling time to remain on the same cold branch as the dense radiatively-efficient circumbinary disk. We therefore turn off cooling in the mini-disk/cavity region, while maintaining the entropy-target cooling prescription in the CBD.

Figure~\ref{fig:hot_rho_2d} illustrates the qualitative change in the inner cavity flow. Initially thin (cold and near ballistic) streams enter the cavity from the inner edge of the circumbinary disk, but after heating up near the black holes they cannot radiate their dissipated orbital energy. The post-shock gas therefore remains hot, expands, and the cavity is filled with filamentary, turbulent material rather than clean ballistic streams. Part of the gas that remains bound to either black hole can still circularize, but it forms geometrically thick mini-disks  (Fig.~\ref{fig:hot_rho_2d}, left panel).

The mass contained in the mini-disks is, on average, smaller by a factor $\simeq 2$ compared to the cold mini-disk simulation, with the remaining supplied material distributed through the cavity or expelled vertically (Fig.~\ref{fig:hot_cold_mass}). The periodicity of the mini-disk mass, set mainly by the CBD feeding cycle, remains similar to that in the efficiently cooled case (Fig.~\ref{fig:hot_cold_mass}, lower panel). The horizon accretion rate, however, decreases on average by a factor of $\approx 3$--$5$, and its variability is suppressed (Fig.~\ref{fig:hot_cold_mass}, upper panel). This reduction in variability reflects the loss of coherent stream impacts. In the cold mini-disk model, streams arriving either from the CBD or from the companion mini-disk remain narrow and nearly ballistic until they strike a mini-disk, producing phase-dependent accretion bursts at the beat frequency. In the uncooled-cavity model, the hot gas expands, mixes, and becomes pressure supported before reaching the holes. The flow coherence is therefore partially scrambled, weakening the direct connection between stream impacts, sloshing, and horizon-scale accretion variability.

The magnetic flux reaching the horizons is only weakly affected by the mini-disk thermodynamics, as shown in the upper panel of Fig.~\ref{fig:hot_cold_bfield}. The hot and cold runs maintain comparable unsigned horizon-threading fluxes, with the hot run differing mainly through shorter-timescale fluctuations. The Poynting luminosity measured in the black-hole frame outside the mini-disks is, however, lower in the hot-cavity run by a factor of $\approx 3$, and its time dependence becomes more intermittent, with shorter and sharper bursts.   

In the hot-cavity run, the {material} captured by the black holes forms geometrically thicker mini-disks (Fig.~\ref{fig:hot_rho_2d}, left panel). The larger scale height provides stronger hydrodynamic confinement near each black hole, reducing the opening angle of the magnetized funnel to $\theta_{\rm jet}\lesssim 20^\circ$, compared with the much wider funnels in the cold mini-disk case (Fig.~\ref{fig:rho_2d}, right panel). At the same time, the hotter sloshing region drives stronger outflows between the black holes. These outflows can strongly perturb the weaker jet, distort the funnel wall, and can partially reprocess or dissipate the electromagnetic energy before it reaches larger radii. Thus, mini-disk thermodynamics affects the jets primarily by changing the funnel geometry, baryon loading, and propagation through the cavity, rather than by strongly changing the total magnetic flux delivered to the horizons. In the efficiently cooled cavity, the same unsigned flux is more coherently organized into a clean, open, magnetically dominated funnel.

\section{Discussion}

\subsection{Comparison with previous work}

    Previous horizon-resolving, radiatively cooled GRMHD simulations have mostly
    focused on smaller separations ($r_{12}\lesssim 20\,M$) and slowly-spinning BHs ($\chi\lesssim 0.6$), where the mini-disks are
    compact, rapidly draining, and close to disruption
    \citep{Bowen2018, bowen2019Quasiperiodicity, Combi2022, gutierrez2021Electromagnetic, Avara2024, Ennoggi2025}. Our simulations extend this regime to rapidly spinning BHs and wider separations $r_{12}=30\,M$, where the mini-disks are more massive and persistent. We nevertheless recover strong beat-frequency variability and inter-mini-disk sloshing, showing that stream impacts remain dynamically important even when the mini-disks behave as persistent reservoirs rather than transient structures. For instance, \cite{Ennoggi2025} showed that the mini-disk masses in the last $\sim 30$ orbits prior to merger drop by a factor of $\sim100$, while the mini-disk masses in our simulations only show a drop of $\sim 3$, consistent with the draining of the CBD.
    
    Hot, geometrically thick circumbinary flows have been studied in several GRMHD simulations of binary black holes in full numerical relativity, e.g., \cite{Gold2014, Paschalidis2021}. These models are appropriate
    for low-accretion systems. It is interesting to compare our hot mini-disk simulation with these globally adiabatic calculations.
    Suppressing cooling inside the cavity in our simulation produces vertically extended mini-disks and suppresses coherent stream-driven variability, consistent with results by \cite{Bright2023}. However, at shorter separations, the transient nature of the mini-disks could still imprint strong variability through their filling-and-depleting cycle \citep{bowen2019Quasiperiodicity} even for hot flows. We note that magnetic transport may operate differently between a globally hot flow and a mixed regime as the one we simulated so this should be further investigated.
    
    Our results are complementary to the low-angular-momentum simulations of 
    \cite{ressler2025}, where gas is supplied isotropically to the binary rather than through a relaxed circumbinary disk. The lack of angular momentum and cavity-mediated accretion suppresses the binary variability in the mass and Poynting flux, see also hydrodynamical calculations in \cite{Cattorini2021} and \cite{DuPont2026} for small and large separation binaries, respectively. More importantly, the rotationally-powered jets in the low-angular momentum case are more intermittent, wobblier, and easily disrupted by their interaction with the surrounding gas. In our case, the jets are in a quasi-steady state, constantly interacting along the rotational axis of the binary.

    Recent simulations of strongly magnetized circumbinary disks show that
    large-scale net flux can qualitatively change the cavity dynamics and mini-disks, leading to magnetically dominated (Rayleigh--Taylor-mediated) circumbinary accretion \citep{Most2024,Wang2025, Most2025} or flux-erupting black holes \citep{Manikantan2025}. Our simulations remain in a non-MAD regime both in the cavity and mini-disks, with dimensionless flux $\langle\phi\rangle\simeq 7-15$ threading the horizons. The magnetized cavity in our simulations contains patches of coherent flux that are otherwise disrupted by the flow and are insufficient to magnetically resist the ram pressure of the streams.

\subsection{Caveats}

The main caveat in our simulations is the treatment of photon cooling. We have assumed that the gas cools towards a target entropy on a Keplerian time-scale which, given the initial conditions of the disk, maintains a geometrically-thin $h/r\sim 0.1$ scale-height \citep{noble2009DIRECT,noble2011RADIATIVE}. Although this captures the internal dissipation due to turbulence in the disk and fixes its geometry, the cooling prescription is ad-hoc and not motivated by microphysics; moreover, the mixed thermodynamical regime where the cooling sink is turned off is also ad-hoc. 

The radiative properties of the accreting system can be very different than predicted in our calculations.  Photon transport effects can also have a major effect on the dynamics for a range of accretion rates as seen in single BH simulations (e.g., \citealt{Jiang2014,Sadowski2016,Mishra2016,Jiang2019,Liska2023,Zhang2025,Zhang2026a,Zhang2026b}). For circumbinary disks, this has been recently investigated by \cite{Tiwari2025, Tiwari2025a}, who found that including radiation transport affects the streams and inner edge of the cavity, possibly affecting the variability of the system. Future simulations with radiation physics resolving the inner binary region will be crucial to understand the highly non-trivial thermodynamic properties of mini-disks and their emission, especially near decoupling where the accretion rate in the cavity drops.

Our simulation initial data, starting from a relaxed CBD, neglect the long-term feedback of the inner binary region to the outer disk. Although this is likely reasonable for our weakly magnetized, equal-mass binary, strongly magnetized CBDs can exhibit secular effects such as flux trapping in the cavity \citep{Most2024}. For our hot cavity simulation, the expelled mini-disk gas can also accumulate and grow. Long-term, full horizon-resolving simulations are needed to assess the influence of these effects.

\section{Conclusion}

We have presented three-dimensional GRMHD simulations of a relaxed,
magnetized circumbinary disk accreting onto an equal-mass binary black hole
with separation $r_{12}=30\,M$ and rapidly spinning, aligned black holes
($\chi=0.9$). We evolved the circumbinary disk into a quasi-steady state
before resolving the horizons and later follow the coupled evolution
of the eccentric cavity, analyzing the properties of persistent mini-disks, magnetic-flux transport, and
dual Poynting-dominated jets. Our main conclusions are as follows.

\begin{itemize}

    \item \emph{Variability of the accretion rate {onto the black holes} is controlled by both the CBD lump and
    inter-mini-disk sloshing.}
    The circumbinary disk develops an eccentric cavity and a strong $m=1$
    overdensity at its inner edge. The lump modulates the mass supply to the
    binary when it passes through its periastron on a period of $\sim 5P_{\rm bin}$. Mini-disk mass and horizon accretion rates show a lower-period modulation near
    $\sim 1.25\,P_{\rm bin}$, produced when the binary motion beats
    against the orbiting lump and when sloshing streams impact the receiving
    mini-disk. 

    \item \emph{The mini-disks reach inflow equilibrium over the evolution but still show strong periodicity.}
    At separation of $r_{12}=30M$, the mini-disks are larger and more massive than in previous near-merger/non-spinning simulations, and they do not fully deplete between accretion episodes. They {do,} however drain faster than expected for a local,
    viscous, geometrically thin $\alpha$-disk. Their accretion cycles are driven
    by a sequence of CBD loading, slosh impact, prompt accretion bursts, and
    relaxation. During the burst phase, low-angular-momentum material from the
    companion mini-disk drives enhanced advective angular-momentum fluxes and
    triggers rapid inflow. During the relaxation phase, Maxwell stresses become important, although advective transport remains a major
    part of the angular-momentum budget.

    \item \emph{Magnetic field is advected through the cavity by laminar streams.}
    The bulk circumbinary disk reaches an MRI-turbulent state with moderate
    magnetization, while the overdense lump remains a relatively high-$\beta$
    region. Interior to the cavity edge, the magnetic field is advected primarily by laminar streams rather than by local turbulence. These streams compress and wind the field, producing a predominantly toroidal structure in the flung-back streams. Near the mini-disk edges, however, the toroidal and poloidal components become comparable, indicating that the field delivered to the black holes is reorganized by stream impacts, shocks, and mini-disk turbulence.

    \item \emph{The mini-disks are strongly magnetized but do not become magnetically arrested.}
    The mini-disks reach plasma beta values of order unity, substantially more
    magnetized than the bulk circumbinary disk due to their relatively low-density. The dimensionless magnetic flux threading the horizons has a time-averaged value $\langle\phi_{\rm BH}\rangle\simeq 7-15$. This remains below the usual  magnetically-arrested disk threshold, $\phi_{\rm BH}\sim 40-60$, and we do not observe horizon-scale flux-eruption events. 
    
    \item \emph{Magnetic flux alternates between the two black holes.}
    The horizon-threading flux is not shared equally at all times. Instead,
    magnetic flux alternates between the two black holes on a timescale of
    $\sim 10$--$15P_{\rm bin}$, longer than the lump
    period. We interpret this as a consequence of the phase drift between the
    eccentric cavity and the binary: successive lump passages preferentially feed
    one black hole for several cycles before the feeding channel reverses. This
    produces long-timescale alternation in both the unsigned horizon flux and
    the dimensionless flux.

    \item \emph{Dual Poynting-dominated jets are produced with an efficiency of $\sim 30\%$ and an on-off state alternating between each BH.}
    The horizon-threading flux powers dual Blandford--Znajek-like jets from the rapidly spinning black holes with a total electromagnetic luminosity reaching an efficiency of $L_{\rm EM}/\langle \dot{M} \rangle\sim 0.3$. The two jets are not equally luminous at all times: the dominant jet switches  between the two black holes and follows the alternating horizon-flux behavior. Thus, the binary produces an ``on--off'' state rather than two identical steady jets.

    \item \emph{The two jets interact continuously and form an extended current sheet.}
    The near-horizon funnels have large opening angles due to the lack of hydrodynamical confinement by the thin mini-disks and thus the two jets
    interact above the binary rotation axis. The toroidal field reverses sign
    across the interface between the funnels, producing an extended and persistent current layer. The two magnetized flux tubes are separated by
    a hotter, baryon-loaded contact layer with $\sigma\sim 0.1$ and weak guide
    field, giving an X-type projected topology favorable to reconnection. This region is a natural site for magnetic dissipation, emission, and possible perturbations of the large-scale jet morphology.

    \item \emph{Mini-disk thermodynamics controls the coherence of accretion
    and jet propagation.}
    When cooling is suppressed inside the cavity, the shocked/turbulent gas remains
    hot, expands vertically, and forms thicker, less massive mini-disks. The
    mini-disk mass decreases by a factor of $\simeq 2$, while the horizon
    accretion rate decreases by a factor of $\sim 3$--$5$ and becomes less
    coherently modulated. The magnetic flux reaching the horizons is only
    weakly affected, but the Poynting luminosity measured outside the mini-disks
    drops by a factor of $\sim 3$. This indicates that the hot cavity changes how the magnetic flux is confined, mass-loaded, and propagated through the cavity. In the efficiently cooled case, the same unsigned flux is more coherently arranged into clean, open, magnetically dominated funnels. 

\end{itemize}

%Overall, our results show that small-separation, rapidly spinning black-hole binaries can produce persistent dual jets even in a non-MAD accretion state.  The electromagnetic variability is regulated by a hierarchy of coupled processes: the CBD lump controls the long-period mass supply, sloshing  governs shorter horizon-scale accretion bursts, magnetic-flux transport determines which black hole powers the dominant jet, and mini-disk thermodynamics controls the coherence of the inner flow and the propagation of Poynting flux through the cavity. Future simulations with radiation transport, explicit resistivity or kinetic reconnection physics, longer horizon-resolved evolution, and different separations and mass ratios will be needed to determine how these signatures evolve toward decoupling and merger.

%\section*{acknowledgments}
\begin{acknowledgments}

We thank our collaborators Michail Chabanov, Elias Most, Geoffrey Ryan, Julian Krolik, Lorenzo Ennoggi, Eduardo Gutierrez, Bart Ripperda and Alexander Philippov for their valuable comments and discussions.

Financial support for the lead author, LC, was provided by the Natural Sciences and Engineering Research Council of Canada (NSERC) through grant DIS-2022-568580. LC also acknowledges support as a CITA National Fellow and the Kavli Institute for Particle Astrophysics and Cosmology (KIPAC) Fellowship. M.C. was supported by a NASA Theory and Computational Astrophysics Network (TCAN) grant (80NSSC24K0100) and NSF awards AST-2009330, AST-1516150, AST-2319326, PHY-2110338, PHY-1707946, PHY-2207920, PHY-2513442, PHY-2409706, and OAC-2411068. 
SMR acknowledges the support of the Natural Sciences and Engineering Research Council of Canada (NSERC), [funding reference number 568580]
Cette recherche a \'et\'e financ\'ee par le Conseil de recherches en sciences naturelles et en g\'enie du Canada (CRSNG), [num\'ero de r\'ef\'erence 568580]. A.J.D. was supported by NASA through the NASA Hubble Fellowship grant No. HST-HF2-51553.001, awarded by the Space Telescope Science Institute, which is operated by the Association of Universities for Research in Astronomy, Inc., for NASA, under contract NAS5-26555.The authors gratefully  acknowledge the computing time made available to them on the high-performance computer “Lise” at the NHR Center NHR@ZIB. This center is jointly supported by the German Federal Ministry of Education
and Research and the state governments participating in the NHR (www.nhr-verein.de/unsere-partner).

Computing resources were provided by SciNet (www.scinethpc.ca), Compute Canada (www.computecanada.ca), the Center for Computational Relativity and Gravitation, and Research Computing at the Rochester Institute of Technology (RIT). Furthermore, M.C. and F.C. acknowledge the Texas Advanced Computing Center (TACC) for access to the Frontera supercomputer through allocations PHY-20010 and AST-20021. 

\end{acknowledgments}

\software{The Einstein Toolkit (\citealt{loffler2012Einstein}; \href{http://einsteintoolkit.org}{http://einsteintoolkit.org}), \texttt{GRMHD\_con2prim} (\citealt{siegel2018Recovery}, \citealt{siegel2018soft}), \texttt{PyCactus} (\citealt{kastaun2021numerical}, \url{https://github.com/wokast/PyCactus}), \texttt{Matplotlib} \citep{hunter2007Matplotlib}, \texttt{NumPy} \citep{harris2020Array}, \texttt{SciPy} \citep{virtanen2020SciPy}, and \texttt{hdf5} \citep{hdf5}. ChatGPT 5.6 (Sol) was used for improving specific parts of the text and physics discussion.}

\newpage 

\appendix

\section{Cooling prescription}
\label{app:cooling}

We apply a local sink term that removes internal energy from gas whose entropy exceeds a prescribed target value. We follow closely the prescription presented originally in \cite{noble2009DIRECT} and used in binary black hole accretion in \cite{noble2012Circumbinary, Bowen2018, Avara2024, Ennoggi2025}. The prescription is intended to maintain a geometrically thin circumbinary disk while allowing shock-heated gas in the cavity to cool on an orbital timescale. The entropy proxy is defined as $K \equiv {P}/{\rho^\Gamma}$, where $\rho$, $P$, and $\Gamma$ are the rest-mass density, pressure, and adiabatic index.  Cooling is applied only above a density threshold, $\rho \geq \rho_{\rm cool}$, and only when $K>K_0$, where $K_0=0.01$ is the target entropy.  The comoving cooling rate per unit volume, $\Lambda$, is applied as
\begin{equation}
\nabla_a T^{ab} = -\Lambda u^b, \quad 
  \Lambda =
  \frac{\rho \epsilon}{t_{\rm cool}}
  \left(
    \frac{K}{K_0}-1
    +
    \left|\frac{K}{K_0}-1\right|
  \right)^q
  {\cal C}({\bf x},t),
  \label{eq:cooling_rate}
\end{equation}
where $\epsilon$ is the specific internal energy, $q$ controls the stiffness of the cooling, and ${\cal C}$ is a spatial switch equal to either zero or one.  Thus $\Lambda=0$ for gas colder than the entropy target. The cooling time is chosen to be proportional to a local Keplerian orbital period,
\begin{equation}
  t_{\rm cool}(M,\chi,r)
  =
  \frac{2\pi \beta}{\Omega_K(M,\chi,r)}, \quad  \Omega_K(M,\chi,r) M
  =
  \frac{1}{(r_{\rm eff}/M)^{3/2}+\chi},
  \qquad
  r_{\rm eff}\equiv \max(r,r_{\rm ISCO}),
  \label{eq:tcool_def}
\end{equation}
i.e., inside the ISCO, the cooling time is frozen to its ISCO value. We use $\beta=1.34$. The cooling time is then assigned by region:
\begin{equation}
t_{\rm cool} =
\begin{cases}
  t_{\rm cool}(M_1,\chi_1,r_1),
  & \bar{r}_1 \leq r_{{\rm md},1}, \\[4pt]
  t_{\rm cool}(M_2,\chi_2,r_2),
  & \bar{r}_2 \leq r_{{\rm md},2}, \\[4pt]
  t_{\rm cool}(M_1+M_2,0,r),
  & r\geq r_{\rm cav}, \\[4pt]
  t_{\rm cool}(M_1+M_2,0,r_{\rm cav}),
  & \text{otherwise}.
\end{cases}
\label{eq:tcool_regions}
\end{equation}
where $r_{\rm md}$ is the mini-disk truncation radius. The last branch corresponds to gas in the cavity between the mini-disks and the circumbinary disk, for which the cooling time is fixed to the orbital time at the inner edge of the circumbinary disk.  In runs where only circumbinary-disk cooling is desired, ${\cal C}=0$ inside the mini-disk regions and ${\cal C}=1$ in the cavity and circumbinary disk. The sink is applied to the conserved energy and momentum variables. In the Valencia $3+1$ formulation, the cooling source terms are
\begin{align}
  \left(\partial_t \tau\right)_{\rm cool}
  &=
  -\alpha \sqrt{\gamma}\, W\,\Lambda,
  \label{eq:tau_cool_source}
  \\
  \left(\partial_t S_i\right)_{\rm cool}
  &=
  -\alpha \sqrt{\gamma}\, W\,\Lambda\, v_i .
  \label{eq:mom_cool_source}
\end{align}
Thus the cooling removes energy and the corresponding advected momentum in the fluid rest frame.  The momentum source is applied explicitly,
\begin{equation}
  S_i^{n+1}
  =
  S_i^n
  -
  \Delta t\,\alpha\sqrt{\gamma}\,W\,\Lambda\,v_i ,
\end{equation}
whereas the energy sink is applied in semi-implicit form to preserve positivity \citep{radice2016Dynamical},
\begin{equation}
  \tau^{n+1}
  =
  \frac{\tau^n}
  {1+\Delta t\,\alpha\sqrt{\gamma}\,W\,\Lambda/\tau^n}.
  \label{eq:semi_implicit_tau}
\end{equation}

\section{Radial histograms for spherically-integrated properties}
\label{app:diagnostics}

To measure properties of the flow around each black hole, such as mass and angular momentum fluxes, we need to construct spherical grids centered on each hole. This can be an expensive operation in post-processing since it requires outputting a lot of 3D data and interpolating onto a new grid for each snapshot.  

We choose instead to use radial histograms on the fly to compute a number of radial-dependent, spherical-averaged properties. Setting a coordinate system around each black hole, we calculate first the volume integrated property $\mathcal{Q}$ on a (lego-like) shell of thickness $\Delta r$ as:
\begin{equation}
    [\mathcal{Q}](\Delta r, r) = \sum^{r+\Delta r}_{r} dV \mathcal{Q},
\end{equation}
where $dV = dx\,dy \,dz$ is the volume of the grid, properly weighted to take into account the local mesh-refinement level, and $\mathcal{Q}$ is a quantity transformed to the BH frame. This operation is performed locally on each MPI processor and then reduced over all processors to get the final volume. Notice that no explicit interpolation is done in this way. We also save the volume of the shell as $[V](\Delta r,r)=    \sum^{r+\Delta r}_{r} dV $. This is done for a set of radii $r_{0} + i\,\Delta r$ logarithmically distributed. 

For sufficiently small $\Delta r$, we would recover a surface-averaged quantity:  
\begin{equation}
   \lim_{\Delta r\rightarrow0} \frac{[\mathcal{Q}](\Delta r, r)}{[V](\Delta r,r)} \approx \frac{\int dA \mathcal{Q}}{\int dA} =\langle \mathcal{Q} \rangle.
\end{equation}

This provides us with cheap, high-speed output to compute fluxes, surface-averaged properties, and volumes around each BH and the center of mass. For instance, to compute the mass flux, we approximate the integral as:
\begin{equation}
4\pi r^2\,\langle {u^r \rho \sqrt{-g}} \rangle \approx \int dA \sqrt{-g} \,\rho u^r = \dot{M}.
\end{equation}

\begin{figure}
    \centering
    \includegraphics[width=1\linewidth]{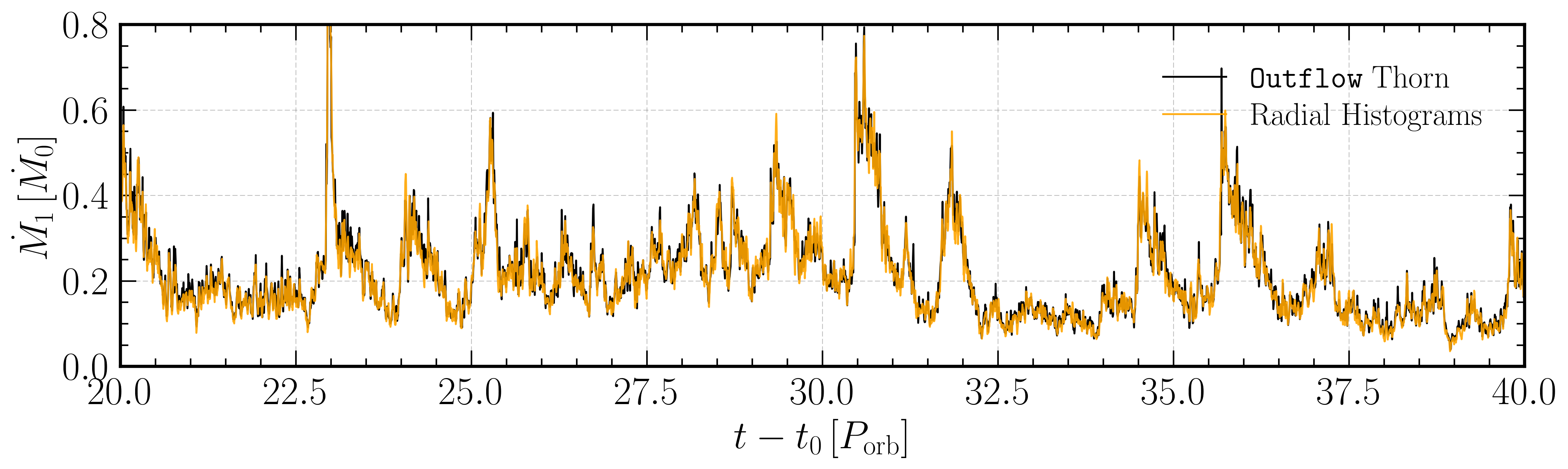}
    \caption{Comparison of accretion rate onto BH$_1$ calculated with a proper interpolation onto the horizon using the \texttt{Outflow} thorn (black) and the radial histograms in the BH frame with the spherical-averaged formalism (yellow).}
    \label{fig:mdot1_comparison}
\end{figure}

This arguably crude approximation works very well when compared to a proper interpolation onto a moving sphere as computed in the \texttt{Outflow} thorn. We show the comparison of the two methods in Fig.~\ref{fig:mdot1_comparison} for the accretion rate of BH$_1$ {and the difference is qualitatively unimportant}.

Similarly, we can define a density-weighted quantity by saving $\rho\mathcal{Q}$ and $\rho$.

\begin{equation}
   \frac{\sum dV \rho\mathcal{Q}}{\sum dV \rho} \approx \langle \mathcal{Q} \rangle_{\rho}.
\end{equation}

Finally, it is straightforward to compute volume integrated quantities like total energy within a radius as $\sum^{r_f}_{r_i} [\mathcal{Q}](\Delta r, r)$. 

\bibliography{astrograv}
\bibliographystyle{aasjournal}

%% This command is needed to show the entire author+affiliation list when
%% the collaboration and author truncation commands are used.  It has to
%% go at the end of the manuscript.
%\allauthors

%% Include this line if you are using the \added, \replaced, \deleted
%% commands to see a summary list of all changes at the end of the article.
%\listofchanges

\end{document}